\RequirePackage{fix-cm}
\documentclass[smallextended]{svjour3}       % onecolumn (second format)
\smartqed  % flush right qed marks, e.g. at end of proof
\usepackage{xspace}
\usepackage{mdframed}
\usepackage{color, colortbl}
\usepackage{array}
\usepackage{hyperref}
\usepackage{graphicx}
\usepackage{multirow}
\usepackage{soul}
\usepackage{lscape}
\usepackage{cleveref}
\usepackage{multirow}
\usepackage{booktabs}
\usepackage{longtable}
\usepackage{caption}
\usepackage{subcaption}
\usepackage{footmisc}
\usepackage[numbers,sort&compress]{natbib}
\usepackage[table,xcdraw]{xcolor}%
\usepackage{titlesec}
\titleformat{\paragraph}
{\normalfont\normalsize\bfseries}{\theparagraph}{1em}{}
\titlespacing*{\paragraph}
{0pt}{3.25ex plus 1ex minus .2ex}{1.5ex plus .2ex}

\usepackage{fancybox}
\newcommand{\hypobox}[1]{

        \begin{center}\noindent\thicklines\setlength{\fboxsep}{6pt}\cornersize{0.2}\ovalbox{

                \begin{minipage}{4.0in}

                        \textit{#1}

                \end{minipage}} 

        \end{center}} 

\mdfdefinestyle{MyFrame}{
 outerlinewidth=0pt,
 skipabove=0pt,
 skipbelow=0pt,
 innertopmargin=5pt,
 innerbottommargin=0pt,
 linewidth=0pt,
 topline=false,
 rightline=false,
 leftline=false,
 innerrightmargin=4pt,
 innerleftmargin=4pt,
 nobreak=false}

\usepackage{amssymb}% http://ctan.org/pkg/amssymb
\usepackage{pifont}% http://ctan.org/pkg/pifont
\newcommand{\cmark}{\textcolor{green}{\ding{51}}}%
\newcommand{\xmark}{\textcolor{red}{\ding{55}}}%

\newcommand{\GH}{{\sc GitHub}\xspace}
\newcommand{\DP}{{\sc Devpost}\xspace}
\newcommand{\WOC}{{\sc World of Code}\xspace}
\newcommand{\RQ}[2]{
\begin{mdframed}[style=MyFrame]\noindent
	\textbf{RQ}$_{#1}$.~\emph{#2}
\end{mdframed}
}

\makeatletter
\newcommand{\linebreakand}{%
  \end{@IEEEauthorhalign}
  \hfill\mbox{}\par
  \mbox{}\hfill\begin{@IEEEauthorhalign}
}
\let\cl@chapter\undefined
\makeatother

\definecolor{Gray}{gray}{0.9}

\newcolumntype{L}[1]{>{\raggedright\let\newline\\\arraybackslash\hspace{0pt}}m{#1}}
\newcommand*{\tableIndent}{\hspace*{0.5cm}}

\usepackage{ifthen}
\usepackage{amssymb}
\DeclareOldFontCommand{\sf}{\normalfont\sffamily}{\mathsf}
\newboolean{showcomments}
\setboolean{showcomments}{true}
\ifthenelse{\boolean{showcomments}}
  {\newcommand{\nb}[2]{
    \fcolorbox{Gray}{yellow}{\bfseries\sffamily\scriptsize#1}
    {\sf\small$\blacktriangleright$\textit{#2}$\blacktriangleleft$}
   }
   
  }
  {\newcommand{\nb}[2]{}
   
  }

\newcommand{\datadoi}{%
  \begingroup\normalfont
  \smash{\href{https://doi.org/10.5281/zenodo.6578707}{\includegraphics[height=1.3\fontcharht\font`\B,trim=0 3 0 0]{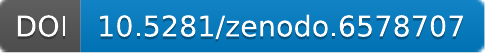}}}%
  \endgroup
}

\begin{document}

\title{One-off Events? An Empirical Study of Hackathon Code Creation and Reuse}
% \subtitle{A quantitative study}


\author{Ahmed Samir Imam Mahmoud \thanks{The First and the Second author contributed equally to this work.} \and
        Tapajit Dey \and 
        Alexander Nolte \and 
        Audris Mockus \and
        James D. Herbsleb
}

\authorrunning{Imam Mahmoud \& Dey et al.} % if too long for running head

\institute{Ahmed Samir Imam Mahmoud \at
              University of Tartu, Estonia \\
              \email{ahmed.imam.mahmoud@ut.ee}
           \and
           Tapajit Dey \at
           Lero---the SFI Research Centre for Software, University of Limerick, Ireland\\
           \email{tapajit.dey@lero.ie}
           \and
           Alexander Nolte \at
University of Tartu, Estonia\\
Carnegie Mellon University, Pittsburgh, PA, USA\\
\email{alexander.nolte@ut.ee}
\and 
Audris Mockus
\at 
University of Tennessee, Knoxville, TN, USA\\
\email{audris@utk.edu}
\and
James D. Herbsleb
\at 
Carnegie Mellon University, Pittsburgh, PA, USA\\
\email{jdh@cs.cmu.edu}
}

\date{Received: date / Accepted: date}

\maketitle

%%%%%%%%%%%%%%%%%%%%%%%%%%%%%%%%%%%%%%%%%
\begin{abstract}
Background: Hackathons have become popular events for teams to collaborate on projects and develop software prototypes. Most existing research focuses on activities during an event with limited attention to the evolution of the hackathon code.
Aim: We aim to understand the evolution of code used in and created during hackathon events, with a particular focus on the code blobs, specifically, how frequently hackathon teams reuse pre-existing code, how much new code they develop, if that code gets reused afterwards, and what factors affect reuse.
Method: We collected information about 22,183 hackathon projects from \DP and obtained related code blobs, authors, project characteristics, original author, code creation time, language, and size information from \WOC. We tracked the reuse of code blobs by identifying all commits containing blobs created during hackathons and identifying all projects that contain those commits. We also conducted a series of surveys in order to gain a deeper understanding of hackathon code evolution that we sent out to hackathon participants whose code was reused, whose code was not reused, and developers who reused some hackathon code.
Result: 9.14\% of the code blobs in hackathon repositories and 8\% of the lines of code (LOC) are created during hackathons and around a third of the hackathon code gets reused in other projects by both blob count and LOC. The number of associated technologies and the number of participants in hackathons increase reuse probability.
Conclusion: The results of our study demonstrates hackathons are not always ``one-off'' events as the common knowledge dictates and it can serve as a starting point for further studies in this area. 

\end{abstract}

%%%%%%%%%%%%%%%%%%%%%%%%%%%%%%%%%%%%%%%%%%%%%
\section{Introduction}
\label{sec:intro}
Hackathons are time-bounded events during which individuals form -- often ad-hoc -- teams and engage in intensive collaboration to complete a project that is of interest to them~\cite{pe2019designing}. They have become a popular form of intense collaboration with the largest collegiate hackathon league alone reporting that their events attract more than 65,000 participants each year\footnote{\url{https://mlh.io/about}}.
The success of hackathons can at least partially be attributed to them being perceived to foster learning~\cite{porras2019code,fowler2016informal,nandi2016hackathons} and community engagement~\cite{nolte2020support,huppenkothen2018hack,taylor2018strategies,moller2014community} and tackle civic, environmental and public health issues~\cite{hope2019hackathons,taylor2018strategies,baccarne2014urban} which led to them consequently being adopted in various domains including
(higher) education~\cite{porras2019code,gama2018hackathon,kienzler2017learning},
(online) communities~\cite{huppenkothen2018hack,taylor2018everybody,busby2016closing,craddock2016brainhack}, 
entrepreneurship~\cite{cobham2017appfest2,nolte2019touched},
corporations~\cite{pe2019designing,nolte2018you,komssi2015hackathons,rosell2014unleashing}, and others. 

Most hackathon projects focus on creating a prototype that can be presented at the end of an event~\cite{medina2020what}. This prototype often takes the form of a piece of software. The creation of software code can, in fact, be considered as one of the main motivations for organizers to run a hackathon event. Scientific and open source communities, in particular, organize such events with the aim of expanding their code base~\cite{pe2019understanding,stoltzfus2017community}. 
It thus appears surprising that the evolution of the code used and developed during a hackathon has not been studied yet, as revealed by a review of existing literature.

In this work, which is an extension of our previous work~\cite{imam2021secret},  we study the evolution of the code used and created by the hackathon team members by focusing on their origin and subsequent reuse. Similar to the previous work, we focus exclusively on the code blobs~\footnote{A blob is a byte string representing a single version of a file, see  \url{https://git-scm.com/book/en/v2/Git-Internals-Git-Objects} for more details} in this study, which is taken as an abstraction for the code itself. It is also worth mentioning that when we talk about code reuse in the context of our study, we are referring to the \textit{exact copying} of a code blob so that the SHA1 value remains the same, i.e. identical code segments with no changes in comments, layouts, and whitespaces.

First, we study from where the code blobs \textit{originate}: While teams will certainly develop original code during a hackathon, it can be expected that they will also utilize existing (open source) code as well as code that they might have created themselves prior to the event.

Second, to understand the impact of hackathon code, i.e. code created during a hackathon event by the hackathon team in the hackathon project repository, we aim to study whether and how the code blobs \textit{propagate} after the event has ended. There are studies on project continuation after an event has ended~\cite{nolte2020what,nolte2018you}. These studies, however, mainly focus on the continuation of a hackathon project in a corporate context~\cite{nolte2018you} and on antecedents of continuous development activity in the same repository that was utilized during the hackathon~\cite{nolte2020what}. The question of where code blobs that has been developed during a hackathon potentially gets reused outside of the context of the original hackathon project has not been sufficiently addressed.

Moreover, we aim to understand what factors might influence the reuse of hackathon code blobs, which can be useful for hackathon organizers and participants to foster the impact of the hackathon projects they organize/participate in. These factors would also be of interest to the open source community in general in order to effectively tap into the potential of hackathons as a source of new software code creation. 

% Finally, we want to examine if the participants' perceived usefulness of the code and the hackathon project itself 

To address the above-mentioned goals, we conducted an archival analysis of the source code utilized and developed in the context of 22,183 hackathon projects that were listed in the hackathon database \DP\footnote{https://devpost.com/}. To track the origin of the code blobs that were used and developed by each hackathon project and study its reuse after an event has ended we used the open source database \WOC~\cite{ma2019world, ma2020world} which allows us to track code (blob) usage between repositories. Overall, we looked at over 8.5M blobs, over 3M of which were code blobs, as identified with the help of the \GH \textit{linguist}~\footnote{\url{https://github.com/github/linguist}} tool. In order to further test the validity of the findings of the data-driven study, we conducted surveys of 178 hackathon participants whose code was reused, 120 hackathon participants whose code was not reused, and 118 OSS developers who reused some code first used in a hackathon.

Our findings indicate that around 9.14\% of the code blobs in hackathon projects are created during an event (8\% if we consider the lines of code), which is significant considering the time and team member constraints. Teams tend to reuse a lot of existing code blobs, primarily in the form of packages/frameworks. Many of the projects we studied focus on front-end technologies -- JavaScript in particular -- which appears reasonable because teams often have to present prototypes at the end of an event, which lends itself to UI design. Most  (45.6\%  overall)  of the hackathon participants indicated (through surveys) that they wrote the code themselves during the hackathon and a number of them also indicated they found the code from the web or wrote it together with another participant. 
Approximately a third of code blobs created during events get reused in other projects, both by the number of blobs and the lines of code, with most being reused in small projects. Most developers who reused some hackathon code (blobs) rated (through the survey) the code to be easy to use. The characteristics of the blobs that were reused, in terms of size  (lines of code)  and whether the code is \textit{template code} (see \cref{sec:rq} for definition) or not, were found to be significantly different from those of the code blobs that were not reused. In most cases, the hackathon code creators did little to foster the reuse of their code and were not aware if it was being reused or not. Moreover, though they expressed satisfaction with the project, they had little intention of continuing to work on it.
The number of associated technologies and the number of participants in a project were found to increase the code blobs' reuse probability, and so did including an Open Source license in the project repository.

In summary, we make the following contributions in the paper: we present an account of code (blob) reuse both by hackathon projects and of the code (blobs) generated during hackathons based on a large-scale study of 22,183 hackathon projects and 1,368,419 projects that reused the hackathon code blobs. We tracked the origins of the code blobs used in hackathon projects, in terms of when it was created and by whom, and also its reuse after an event. We gained further insight into the evolution of the code (blobs) by conducting a series of surveys and also identified a number of project characteristics that can affect hackathon code blob reuse. 

The replication package for our study is available at~\cite{repPackage}, and the processed dataset containing the SHA1 hash values for the final list of blobs under consideration, the ID of the hackathon (in \DP) the blob is associated with, the \GH project/repository name (in \texttt{owner/repository} format), and the various attributes of the blobs we identified, viz., when the blob was first created and by whom, whether the blob was reused afterwards, and if so, in what type of project, and if the blob is a ``template code'' blob or not, is available at \datadoi\cite{ahmed_samir_imam_mahmoud_2022_6578708}.

\section{Research Questions}
\label{sec:rq}
As mentioned in~\cref{sec:intro}, the goal of this study is to understand the evolution of hackathon code blobs and identify factors that affect code blob reuse for these projects. 

\vspace{10pt}
% \textbf{Code Origin:}
\noindent Our \textbf{first research question} thus addresses the origin of hackathon code blobs:
\RQ{1}{Where does the code blobs used in hackathon projects originate from?}
%\vsapce{-8pt}
Delving deeper into this question, we aim to understand how much of the code blobs used in a hackathon project was actually created \textit{before} the event and reused in the project, how much of the code blobs were developed \textit{during} the hackathon, and, since the projects sometimes continue even after the official end date of the hackathon, how much of the code blobs were created \textit{after} the event. This leads us to the sub-question:
\RQ{1a}{When were the code blobs created?} 
We also aim to understand how much of the code blobs in a hackathon project repository are created by one of the participants, how frequently they reused code blobs created by someone they worked with earlier, and how much of code was created by someone else, leading us to the sub-question:
\RQ{1b}{Who were the original creators of the code blobs?}
%\vsapce{-8pt}
A related question that can lead to a deeper understanding of the hackathon projects and the code is related to the source of the code blobs - was it written by the project participants during the hackathon, did they reuse code blobs from some other projects/from online communities like StackOverflow, or did the code blobs have a different origin? This leads us to the following sub-question:
\RQ{1c}{Where did the code blobs originate from?}
We also wanted to examine what are the programming languages predominantly used in various hackathon projects and the distribution of languages in the code created before, during, and after the projects:
\RQ{1d}{What are the languages of hackathon code created before, during, and after the event?}
In addition, to quantify ``how much'' code in various hackathon projects was created before, during, or after the hackathon, we also want to measure the sizes of the code blobs, leading us to the sub-question:
\RQ{1e}{What are the sizes of code blobs created before, during, and after the hackathons?}
Finally, we focused on the fact that a lot of developers often use some boilerplate code to initialize their repositories and a good chunk of code in many repositories is actually created by various automated tools, scripts, or are part of some libraries or frameworks. The code blobs related to to such code is generally different from code written/reused by developers because these code blobs are meant to be used ``as-is'', in contrast to some code blob a developer in a project might want to reuse ``as-is'' by their conscious decision. We call such code (blobs) ``template code'', which we define as: \textit{``ready-to-use'' code that is generally intended to be used ``as-is'', e.g. code (blobs) that is part of a library/framework, or is a commonly used boilerplate code, e.g. boilerplate code (blobs) used while initializing a project or blobs representing contents of placeholder files.}
We decided to differentiate such ``template code'' from code written/reused by human developers because ``template code'' is developed with the intention of being used ``as-is'' while non-template code is not, leading us to the sub-question:
\RQ{1f}{What fraction of code blobs created before, during, or after the hackathons can be categorized as template code?}

\vspace{10pt}

Our \textbf{second research question} focuses on the aspect of hackathon code (blob) reuse. As noted in~\cref{sec:intro}, existing studies do not address the question of whether and where hackathon code blobs get reused after an event has ended. However, knowing the answer to this question would be crucial for understanding the impact of hackathons on the larger open source community. Some might perceive hackathons as one-off events where people gather and create some code that is never used again, while in fact they might have an impact on the wider scene of software development and create something of value that transcends individual events. This leads us to also ask the following second research question:
\RQ{2}{What happens to hackathon code (blobs) after the event?}
To address this question, we started by looking at what fraction of the hackathon code blobs get reused and the characteristics of the reused code blobs in terms of size, purpose (whether it is a template code or not), programming language the code was written in, and ease-of-use (based on responses from developers who reused the code blobs). We focused on these factors because previous studies (see \cref{sec:lit:reuse}) suggested that these are important characteristics of the code blobs that gets reused. So our first sub-question is:
\RQ{2a}{How much hackathon code (blobs) gets reused and what are the distributions for size, purpose, language, and ease-of-use of the reused code blobs?}
We also wanted to know if the characteristics of the reused code blobs, as mentioned in the last sub-question, differ significantly from that of the code blobs which didn't get reused, leading to the second sub-question:
\RQ{2b}{Are the characteristics of the reused hackathon code blobs significantly different from those of the blobs that do not get reused?}
In terms of assessing the impact of the code blobs that were reused, arguably, some code blob that is reused in a large project reaches and ultimately impacts more people, and thus can be perceived to have more impact on and be more useful to the software development community overall, than code blobs reused in a small project, so we wanted to explore how frequently the hackathon code blobs get reused in small/medium/large OSS projects:
\RQ{2c}{What are the sizes of the projects that reuse hackathon code blobs?}
Next, we wanted to explore if the hackathon project members who created the code somehow foster code reuse, examining if they themselves reused the code (blobs) or shared the code afterwards:
\RQ{2d}{How commonly do the hackathon code owners foster code (blob) reuse by actively reusing/sharing the code/using a license in the repository that allows code reuse?}
We also wanted to examine if the hackathon code owners are aware that some of the code they created during a hackathon might be getting reused and if they feel that their code is useful for others, which ties to the aspects of transparency and perception of the participants:
\RQ{2e}{Are the hackathon code authors aware of code reuse/feel their code might be useful to others?}
Finally, though not a form of direct code reuse, we wanted to know if the ideas developed in the hackathon projects are ever reused because this is another form of continuation of the hackathon project which is often hard to capture from purely data-driven studies:
\RQ{2f}{Do the ideas that arose from a hackathon project get reused?}

\vspace{10pt}

Our \textbf{third research question} focuses on understanding how different characteristics of a hackathon project can influence the probability of hackathon code (blob) reuse.~\footnote{Here we are interested in discovering if the independent variables are able to statistically explain the dependent variable, not causal inference.} While code reuse in Open Source Software is a topic of much interest, there are only a few studies covering this topic. Moreover, existing studies, e.g.~\cite{haefliger2008code,kawamitsu2014identifying,feitosa2020code,von2005knowledge} only focus on between 10 and a few hundred projects. For this study, we examined 22,183 hackathon projects, which makes it reasonable to assume that insights from this study -- despite them being drawn from hackathon projects only and being focused on reuse at the blob level (i.e. exact copy-paste of code) -- would add to the existing knowledge about code reuse in general. Thus, we present this question as:
\RQ{3}{How do certain project characteristics influence hackathon code (blob) reuse?}
Related to this research question, we formed the following hypotheses that focus on aspects that can reasonably be expected to foster code (blob) reuse:\\
\noindent\textbf{H1 Familiarity:} Projects that are attempted by larger teams will have a higher chance of their code being reused, simply because more people are familiar with the code. Moreover, hackathon events that are co-located offer participants more possibilities for interaction which can contribute to a better understanding of each other's code, higher code quality, and consequently foster code reuse.\\
\noindent\textbf{H2 Prolificness:} Code from projects involving many different technologies is more likely to be reused, since: (a) they tend to have more general-purpose code than more focused projects, which affects code reuse as discussed by Mockus~\cite{mockus2007large}, and (b) they have a cross-language appeal, opening more possibility for reuse. Similarly, projects with more amount of code created before and during the event (we can not use the code created after the event, for preventing data leakage) should have a higher chance of code reuse by virtue of simply having more code.\\
\noindent\textbf{H3 Composition:} The project composition, i.e. how many blobs in a project are actually code blobs, and how many are non-code blobs, e.g., data, documentation, or others could be another factor that might influence code (blob) reuse. This relationship is likely to be non-linear though, e.g., since we are considering code reuse, a higher percentage of code in the project should increase the probability of reuse, but only up to a certain point, since code from a repository containing only code and no documentation is not very likely to be reused.\\
\noindent\textbf{H4 Permission:} Last but not least, we also focus on if the code (blob) is actually shared with an Open Source License which allows it to be reused because, if the hackathon project repository is not licensed to be reused, the chance of that code (blob) being used elsewhere should be very low.

\section{Background}
\label{sec:lit}
In this section we will situate our work in the context of prior research on hackathon code (\cref{sec:lit:hack}) before discussing existing studies on code reuse (\cref{sec:lit:reuse}).

\subsection{Research on hackathon code}
\label{sec:lit:hack}
The rise in popularity of hackathon events has led to an increased interest to study them~\cite{falk202010}. Current research however mainly focuses on the event itself studying how to attract participants~\cite{taylor2018everybody,hou2017hacking}, how to engage diverse audiences~\cite{paganini2020engaging,hope2019hackathons,filippova2017diversity}, how to integrate newcomers~\cite{nolte2020support}, how teams self-organize~\cite{trainer2016hackathon} and how to run hackathons in specific contexts~\cite{moller2014community,pe2019designing,porras2019code}. These studies acknowledge the project that teams work on as an important aspect. The question of where the software code that teams utilize for their project comes from and where it potentially gets reused after an event has not been a strong focus though.

There are also studies that focus on the continuation of software projects after an event has ended~\cite{lapp20072006,cobham2017appfest2,nolte2018you,ciaghi2016hacking}. These studies however mainly discuss how activities of a team during, before, and after a hackathon can foster project continuation~\cite{nolte2018you}, how hackathon projects fit to existing projects~\cite{lapp20072006}, and the influence of involving stakeholders when planning a hackathon project on its continuation~\cite{cobham2017appfest2,ciaghi2016hacking}. They do not specifically focus on the code that is being developed as part of a hackathon project.

Few studies have also considered the code that teams develop during a hackathon~\cite{nolte2020what,busby2016closing}. These studies however mainly focus on code availability after an event~\cite{busby2016closing} or on how activity before and after an event within the same repository that a team utilized during the hackathon can affect reuse~\cite{nolte2020what}. The question of whether and to what extend teams utilize existing code and whether and where the code that they develop during a hackathon gets reused aside from this specific repository has not been addressed.

\subsection{Code reuse}
\label{sec:lit:reuse}
Code reuse has been a topic of interest and is generally perceived to foster developer effectiveness, efficiency, and reduce development costs~\cite{sojer2010code,haefliger2008code,feitosa2020code}. Existing work so far mainly focuses on the relationship between certain developer traits~\cite{haefliger2008code,sojer2010code,von2005knowledge} and team and project characteristics such as team size, developer experience, and project size and code reuse~\cite{abdalkareem2017code}. Moreover, the aforementioned findings are mainly based on surveys among developers, thus covering their perception rather than actual reuse behavior. In contrast, we aim to study actual code reuse behavior.

There is also existing work that focuses on studying the reuse of the code itself. These, however, are often small scale studies of a  few projects~\cite{kawamitsu2014identifying,xu2020reinventing} focusing on aspects such as automatically tracking reuse between two projects~\cite{kawamitsu2014identifying} and identifying reasons why developers might choose reuse over re-implementation~\cite{xu2020reinventing}. In contrast, our aim is to study how the code created during a hackathon evolves i.e. where it comes from and whether and where it gets reused.

Large scale studies on code reuse have been scarce. The few existing studies often focus on code dependencies~\cite{german2007using} or on technical dept induced based on reuse~\cite{feitosa2020code} which are both not a strong focus for us because our aim is rather to study where hackathon code gets reused. There are studies that discuss the reuse of code on a larger scale~\cite{mockus2007large} and showed that it is mainly code from large established open source projects that get reused, while we aim to study reuse of code that has been developed by a small group of people during a short-term intensive coding event.

This work is an extension of our previous work~\cite{imam2021secret,imam2021tracking} where we conducted a study on the evolution of hackathon-related code, specifically, how much the hackathon teams rely on pre-existing code and how much new code they develop during a hackathon. Moreover, we also focused on understanding if and where that code (blob) gets reused and which project-related aspects predicate code (blob) reuse. However, it did not differentiate between template code that is intended to be reused ``as-is'' and code that was written by developers with no such intention. Moreover, it only looked at the number of blobs when considering ``how much'' code was reused, not accounting for the sizes of the code blobs, and also did not investigate the effect of having an Open Source license in the hackathon repository on code (blob) reuse. It did not consider the perception of individuals that created and reused hackathon code either. This study, therefore, makes a novel contribution by providing a more in-depth examination of the creation and reuse of hackathon-related code blobs through additional quantitative and qualitative analysis.

\section{Methodology}
\label{sec:method}

In this section, we describe the methodology followed in this study. As mentioned in \cref{sec:intro}, we conducted a data-driven study using the \DP and the \WOC platforms and also a series of three surveys for addressing our research questions. 

\subsection{Data-Driven Study:}
The data-driven study we conducted involved studying actual hackathon projects using the \DP and \WOC platforms. This study was used for answering \textit{RQs 1a, 1b, 1d, 1e, 2e, and 3} completely and also for partially answering \textit{RQs 2a and 2b}.  
\subsubsection{Data Sources}
While hackathon events have risen in popularity in the recent past, many of them remain ad-hoc events, and thus data about those events is not stored in an organized fashion. 
However, \DP is a popular hackathon database that is used by corporations, universities, civic engagement groups and others to advertise events and attract participants. It contains data about hackathons including hackathon locations, dates, prizes and information about teams and their projects including the project's \GH repositories. Organizers curate the information about hackathons and participants indicate which hackathons they participated in, which teams they were part of and which projects they worked on. \DP does not conduct accuracy checks. 

However, \DP does not contain all the information required for answering our research questions. We thus leveraged the \WOC dataset for gathering additional information about projects, authors, and code blobs. \WOC is a prototype of an updatable and expandable infrastructure to support research and tools that rely on version control data from the entirety of open source projects that use Git. 
It contains information about OSS projects, developers (authors), commits, code blobs, file names, and more. \WOC provides maps of the relationships between these entities, which is useful in gathering all relevant information required for this study.
We used version S of the dataset for the analysis described in this paper which contains repositories identified until Aug 28, 2020.
~\footnote{For further details, please check:\url{https://bitbucket.org/swsc/overview/src/master/}}

%\vsapce{-5pt}
\subsubsection{Data Collection and Cleaning}
\begin{figure}[!t]
    \centering
    \includegraphics[width=\linewidth]{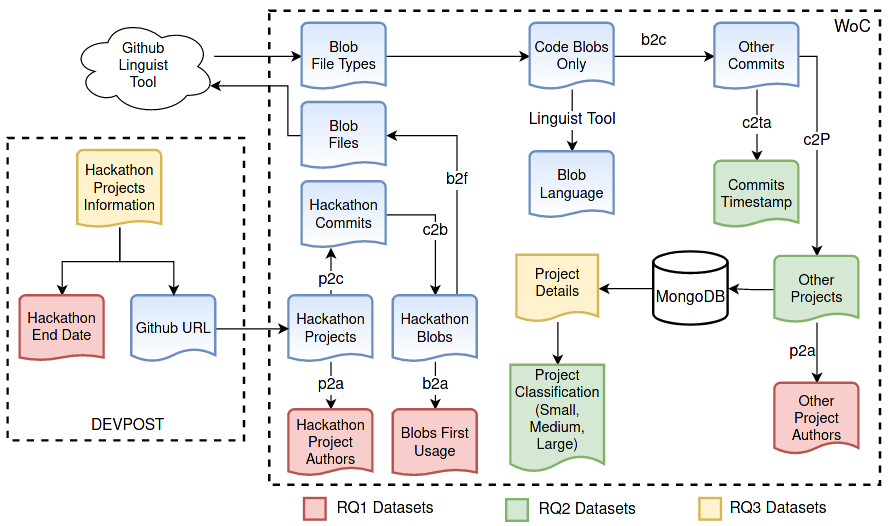}
    \caption{\textbf{Data Collection Workflow: Highlighting the different data sources used and the process of gathering the required information from them, and the data used in answering our research questions}}
    \label{fig:DataCollection}
    %\vsapce{-15pt}
\end{figure}

Here we describe how we collected the data required for answering our research questions, along with the details of all the filtering we introduced. An overview of the approach is shown in~\cref{fig:DataCollection}, which also highlights the different data sources and what data was used for answering each research question.

\paragraph{Selecting appropriate hackathon projects for the study}
We started by collecting information about 60,479 hackathon projects from \DP. Since the project ID used in \DP is different from the project names in \WOC, in order to link these hackathon projects to the corresponding projects in \WOC, we looked at the corresponding \GH URLs, which could be easily mapped to the project names used in \WOC, where the project names are stored as \\\texttt{GitHubUserName\_RepoName}. After filtering out the projects without a \GH URL, we ended up with 23,522 projects. While trying to match these projects with the corresponding ones in \WOC, we were not able to match 1,339 projects, which might have been deleted or had their names changed afterward. Thus, we ended up with 22,183 projects for further analysis.

\paragraph{Characteristics of the chosen hackathon projects}

\begin{figure}[!t]
    \centering
    \includegraphics[width=\linewidth]{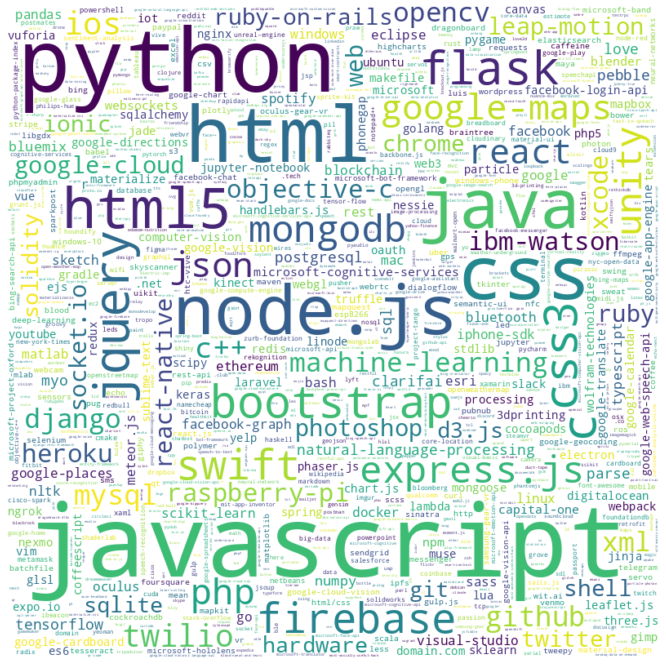}
    \caption{\textbf{Common technologies associated with the chosen hackathon projects}}
    \label{fig:topics}
    %\vsapce{-15pt}
\end{figure}
\begin{figure}[!t]
    \centering
    \includegraphics[width=\linewidth]{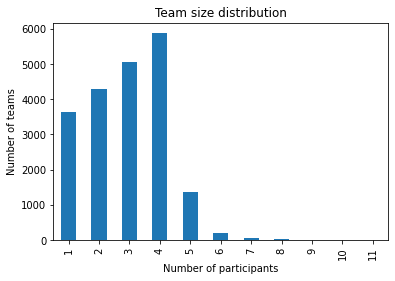}
    \caption{\textbf{Distribution of team sizes for the hackathon projects}}
    \label{fig:teamsize}
    %\vsapce{-15pt}
\end{figure}
\begin{figure}[!t]
    \centering
    \includegraphics[width=\linewidth]{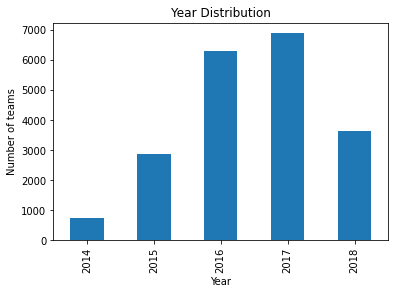}
    \caption{\textbf{Distribution of when the hackathons under consideration took place}}
    \label{fig:year}
    %\vsapce{-15pt}
\end{figure}

To understand the characteristics of the chosen hackathon projects, we decided to look at the various technologies associated with the hackathons, the distributions of team sizes for the hackathon teams, and when the hackathons took place, which are shown in ~\cref{fig:topics},~\cref{fig:teamsize}, and ~\cref{fig:year} respectively. We can see that JavaScript, Python, Java are among the more popular technologies used in the hackathons under consideration. Most of the hackathons tend to have up to 4 members and majority of the hackathons under consideration took place in 2016 and 2017.

\paragraph{Gathering the contents (blobs) of the project}
Our first step was to identify all code blobs used in the hackathon projects. \WOC does not have a direct map between projects and blobs, so we started by collecting commits for all hackathon projects (repositories) using the project-to-commit (\textit{p2c}) map in \WOC (in the \WOC terminology, projects essentially refer to repositories), which covers all commits for each hackathon project.
For the 22,183 hackathon projects, we collected 1,659,435 commits generated in the hackathon repository. 
Then, we gathered all the blobs associated with these commits using the commit-to-blob (\textit{c2b}) map in \WOC, which yielded 8,501,735 blobs, which are all the blobs associated with the hackathon projects under consideration.

\paragraph{Filtering to only select code blobs}
The hackathon project repositories, like most other OSS project repositories, have more than just code in them --- they also contain images, data, documentation, etc. Since our aim in this project is the identification of the reuse of ``code", we decided to filter the blobs to only have the ones related to ``code". In order to achieve that, we looked at the filenames for each of the blobs in the project (since blobs only store the contents of a file, not the file name) using the blob-to-filename (\textit{b2f}) maps in \WOC. After that, we used the \textit{linguist} tool from \GH to find out the file types. The \textit{linguist} tool classifies files into types ``data, programming, markup, prose'', with files of type ``programming'' being what we are focusing on in this study. Additionally, we marked all the files that are not classified by the tool as files of type ``Other'', with the presumption that they do not contain any code. Therefore, we focused only on the blobs whose corresponding files are classified as ``programming'' files by the \textit{linguist} tool, which reduced the number of blobs under consideration to 3,079,487.

\paragraph{Gathering data required to identify the origins of hackathon code (blobs)}
\label{sss:rq1}
To address our first research question, we needed information about the first commits associated with each of the 3,079,487 blobs under consideration. Fortunately, it is possible to get information about the first commit that introduced each blob using \WOC. We extracted the \textit{author} of that first commit, along with the \textit{timestamp}, which would be useful in identifying when the blob was first created. We have the end date for each of the hackathon events from \DP, however, it does not include any information about the start date of the hackathon. We consider the start of a hackathon 72 hours before the end date. This assumption appears reasonable since hackathons are commonly hosted over a period of 48 which are often distributed over three days~\cite{Nolte2020HowTO,cobham2017appfest2}. We also conducted a manual investigation of 73 randomly selected hackathons and found that only 2 projects (2.7\%) were longer than 3 days, which empirically suggested that our assumption would be valid for most of the hackathons under consideration.
Under this assumption, we have the start and end dates of each of the hackathon events, and we used that information to identify if a blob used in a hackathon project was created before, during, or after the hackathon event.

We can identify the first commit that introduced the hackathon blobs under consideration, the \textit{author} of that commit, and all of the developers who have been a part of the hackathon project~\footnote{We used the approach outlined by~\cite{fry2020idres} for author ID disambiguation to merge all of the different IDs belonging to one developer together, which is a common occurrence, as discussed in~\cite{Dey2020RepresentationOD}} using \WOC. With this data, we can determine if the blob was first created by a member of the hackathon team or someone else. In order to dig further and understand if the blob was created in another project a member of the hackathon project participated in, we used \WOC to identify the project associated with the first commit for each blob under consideration, identified all developers of that project, and checked if any of them are members of the team that created the hackathon project under consideration. This lets us identify if the blob was created by (a) a developer who is a participant of the hackathon project (thus, they are creating the code/reusing what they had created earlier), (b) a developer who was part of a project one of the participants of the hackathon project also contributed to (which might suggest that they are familiar with the code, which might have influenced the reuse of that code (blob) in the hackathon project), or (c) someone else who has not contributed to any of the projects the hackathon project developers previously contributed to (which would suggest a lack of direct familiarity with the code from the hackathon participants' perspective). 

\paragraph{Code Size Calculation}\label{sss:loc}
In order to quantify ``how much'' code in various hackathon projects was created before, during, or after the hackathon, we focused on calculating the file size. Measuring file size is straightforward and, arguably, reflects the amount of effort that went into creating a file. One of the most common ways to measure file size is to calculate the number of lines of code (LOC) in the file. However, simply counting the lines is often not enough since code files often contain blank lines and comments along with the actual code, and, arguably, they do not represent the same amount of coding effort. Therefore, in order to get a more accurate estimate of the file sizes, which would, in turn, represent the effort that went into creating the code, we decided to only measure the actual lines of  ``code'' and exclude the blanks and commented lines from consideration.

While there are several tools that can be used to get such information, one of the common tools is \textit{cloc}~\footnote{\url{https://github.com/AlDanial/cloc}} which is an open-source command-line tool that can measure the lines of code for several programming languages. The cloc tool uses the file name to be able to determine the target programming language and then will use that as input to calculate the file statistics (code, comment, and blank lines). The cloc tool can also work by specifying the language directly instead of the file name. The tool reads the blob content and calculates how many code lines, commented lines, and blank lines are in the file. 

We started by identifying the file names for each blob hash using \textit{b2f} map in \WOC. We noticed that the \textit{cloc} tool was not able to identify the language directly from file names for all of the collected blobs, thus we decided to use the \GH Linguist tool to get the language of each blob based on the collected file names. We also collected the content of each blob using \WOC and using the collected data, we were able to run the \textit{cloc} tool with the language and file content as input to identify code size for each blob.  

\paragraph{Template Code Detection}\label{sss:template}
We decided to differentiate between code that is provided as a part of a library/framework/boilerplate template/generated by a tool with the intention of being reused ``as-is'' and code that is written by a developer without such intention because we expect them to have very different characteristics in terms of where they originate from, where they are reused, and also how much actual ``effort" goes into adding this code into a project. However, since we are not limiting our analysis to any one specific language, there is no general way to differentiate between the two with absolute certainty. Therefore, we decided to use a set of three heuristics, formulated by trial-and-error, for detecting template code blobs, described below:

\noindent\textbf{Rule 1: Depth} - The filename associated with the blob has a depth of 3 or more relative to the parent directory of the project.\\
\noindent \textit{Example:} A file located at \texttt{work/opt/tmp.py} is located at a depth of 3 (the file is located in the \texttt{opt} directory which is inside the \texttt{work} directory relative to the parent directory of the project).\\
\noindent \textit{Rationale:} Based on manual empirical observation of 100 different template files selected randomly from various hackathon projects in different languages, only 3 were found to be located at a depth of 2 and 0 at a depth of 1 (the parent directory of the project). 

\noindent\textbf{Rule 2: Similar filename} - If a blob is associated with $n$ filenames in total and $u$ unique filenames, then we expect ratio  $u/n$ to be less than or equal to $0.33$ (if an example blob has these file names associated with it: $[A, A, A, B, B]$, we would have $n = 5$ and $u = 2$).\\
\noindent \textit{Example:} Using the \textit{b2f} (blob to filename) map of \WOC, an example blob was found to be associated with 15 filenames (relative to different projects where the blob was used), and only 2 unique filenames. The ratio of $u/n$ in this case was $0.1333$, so this rule would mark that blob as template code.\\
\noindent \textit{Rationale:} The template files are often reused in various projects \textit{as-is} with the same filename (since they are often a part of some standard library/framework), so we would expect the value of $n$ to be relatively large (because of reuse) and the value of $u$ to be relatively small (because code is reused \textit{as-is}). Based on a manual empirical observation of 100 different template files selected randomly from various hackathon projects in different languages, all of them satisfied the condition $u/n <= 0.33$.

\noindent\textbf{Rule 3: Keywords} - Have one of these keywords in the filename path - \texttt{plugins, frameworks, site-packages, node\_modules, vendor}.\\
\noindent \textit{Example:} One of the filename associated with the blob\\ \texttt{1ff93087f56aeeacece3a3ae4ccf0ba7dff6b8a9} is\\ \texttt{venv/lib/python2.7/site-packages/django/contrib/auth/\_\_init\_\_.pyc}. Since the filename contains the keyword \texttt{site-packages}, it would be marked as a template file according to this rule.\\
\noindent \textit{Rationale:} Based on a manual empirical observation of 100 different template files selected randomly from various hackathon projects in different languages, 83 were found to contain the above-mentioned keywords in at least one of the filenames associated with the blob. This rule addresses the reuse of blobs related to modules/libraries/package files, which fits our definition of ``template code''.

We used these three heuristic rules for detecting template code, which were formulated by trial-and-error, and a different set of 100 code blobs were selected randomly in each case.  Finally, \textbf{all code blobs that satisfied at least 2 of the 3 rules mentioned above were marked as template code for our analysis.}

In order to check the effectiveness of our proposed heuristics in detecting template code, we manually checked 100 randomly selected blobs marked as ``template code'' by our overall heuristics (different from the template code blobs used for formulating the three heuristics rules) and 100 more randomly selected blobs marked to be non-template code by examining the filenames, the blob contents, and the first commits that introduced each blob to discern whether it actually is template code or not. The confusion matrix related to the evaluation is presented in~\cref{t:heuristics}. The associated values of precision, recall (sensitivity), specificity, accuracy, and F1-score are $0.96, 0.79, 0.95, 0.85,$ and $0.86$ respectively. We found that our heuristics are effectively tagging blobs associated with widely used package files, however, it has the problem of wrongly tagging some non-template code blobs as template ones particularly in large projects because the filename might not be that uncommon and they tend to have a sufficiently deep directory structure.
We tested a few other ideas than the ones described above, e.g., identifying ``common modules'' by the number of occurrences with the assumption that ``template code'' blobs are more common than the rest but they didn't affect the overall accuracy of our heuristics too much.

% \usepackage{graphicx}
\begin{table}[!ht]
\caption{Confusion Matrix showing the effectiveness of our heuristics in identifying template code.}
\label{t:heuristics}
\resizebox{0.8\textwidth}{!}{%
\begin{tabular}{lrr}
                   & True Positive & True Negative \\
Predicted Positive & 96            & 4             \\
Predicted Negative & 26            & 74           
\end{tabular}%
}
\end{table}

\paragraph{Gathering data to identify hackathon code blob reuse}
\label{sss:rq2}
Our second research question focuses on the reuse of \textit{hackathon code}, which, per our definition (see ~\cref{sec:intro}), refers to the blobs created during the hackathon event by one of the members of the hackathon team. Therefore, to address this question, we utilized the results of our earlier analysis in order to focus only on the code blobs which satisfy the following two conditions: (a) The blob was first introduced during the hackathon event and (b) the blob was created by one of the hackathon project developers. After identifying 581,579 blobs that met these conditions, we collected all commits containing these blobs from \WOC using the blob-to-commit (\textit{b2c}) map, and we collected the projects where these commits are used using the commit-to-project (\textit{c2p}) map. \WOC has the option of returning only the most central repositories associated to each commit, excluding the forked ones (based on the work published in~\cite{forkrepo}), and we used that feature to focus only on the repositories that first introduced these blobs, and excluded the ones that were forked off of that repository later, since most forks are created just to submit a pull request and counting such forks would lead to double-counting of code blob reuse.

% For the purpose of this study, we focused only on the blob reuse scenario within 2 years of the end date of the corresponding hackathon event, since we believe that any blob reuse that happened after 2 years isn't likely impacted by the hackathon event itself.\ai{please verify}

In addition to understanding how the blobs get reused, we also wanted to understand if they are reused in very small projects, or if larger projects also reuse these blobs. So, we needed a way to classify the projects into different categories. We focused on two different project characteristics for the purpose of such classification: the number of developers who contributed to that project, and the number of \textit{stars} it has on \GH, a measure available from a database (MongoDB) associated with \WOC.

Both the number of developers and \textit{stars} are quintessential measures of project size and popularity and were found to have a low correlation (Spearman Correlation: 0.26), so we decided to use both measures.
Instead of manually classifying the projects using these variables using arbitrary thresholds, we decided to use \textit{Hartemink's pairwise mutual information based discretization method}~\cite{hartemink2001principled}, which was applied to a dataset with log-transformed values of the number of stars and developers for projects, to classify them into three categories: Small, Medium, and Large. We found different thresholds for the number of developers and \textit{stars} (for no. of developers, $>2 \rightarrow$ Medium projects and $>6 \rightarrow$ Large; for stars, $>1 \rightarrow$ Medium and $>14 \rightarrow$ Large), and classified a project as ``Large" if it is classified as such by either the number of developers or the number of \textit{stars}, and used a similar approach for classifying them as ``Medium''. The remaining projects were classified as ``Small''. Overall, we identified 1,368,419 projects that reused at least one of the 581,579 blobs, and using our classification, 1,220,114 (89.2\%) projects were classified as ``Small'', 116,177 (8.5\%) as ``Medium'', and 32,128 (2.3\%) as ``Large''. 

\paragraph{Discovering the Licenses associated with the hackathon projects}
One of the main factors that could affect code blob reuse is the license associated with the hackathon project (H4 for RQ3 as presented in ~\cref{sec:rq}). In order to assess the validity of that claim, we had to collect and identify the type of license associated with the hackathon projects. 

Although \WOC provides a tremendous amount of information about open source projects, it doesn't include the license details for projects. In order to be able to collect the license details, we used \GH REST API to generate calls for collecting license information for the projects under consideration. We used the list of hackathon projects as input and called the license REST API in \GH on each of them. We found that, unsurprisingly, some projects don't have a license at all and some other projects got deleted, merged or renamed. We handled such cases and marked projects with no license as \textit{'NOLICENSE'} and projects which are we could not find in \GH as \textit{'NOTFOUND'}.

We ran the license collection on 27,419 projects. We have more projects to consider here since this was done in the second phase of the study and the number of projects associated with the hackathon blobs increased within that time. 5,910 of the projects were not found in \GH, and 17,742 projects were found to not have a license. The license type used most frequently is "MIT License" with 2,301 projects, followed by "Apache License 2.0" license with 441 projects.

\paragraph{Collecting Data for identifying the factors that affect\\ hackathon code blob reuse}
\label{sss:rq3-data}

\begin{table}[t]
\caption{Description of variables used for addressing RQ3. For the Binary variable, no. of TRUE/FALSE cases are shown}
\label{t:variables}
\resizebox{\textwidth}{!}{%
\begin{tabular}{@{}llp{3.2cm}p{1.5cm}rr@{}}
\toprule
\multirow{2}{*}{\textbf{Hypothesis}} & \multirow{2}{*}{\textbf{Variable}} & \textbf{Variable}  & \multirow{2}{*}{\textbf{Source}} & \textbf{Value Range}  & \multirow{2}{*}{\textbf{Median}} \\
& & \textbf{Description} & & \textbf{(min-max)} & \\\midrule
 & no.Participant & Number of Participants in the hackathon. & \DP & 2 - 10 & 3 \\
\multirow{-2}{*}{H1: Familiarity} & is.colocated & hackathon is held in single or multi location. & \DP & \multicolumn{2}{p{3cm}}{\cellcolor[HTML]{FFFC9E}TRUE: 12,445 (97\%) FALSE:436 (3\%)} \\\hline
 & no.Technology & Number of different technologies the hackathon is related to & \DP & 1 - 40 & 5 \\
 & Before & Number of blobs in the hackathon project repo that were created before the event. & \WOC & 0 - 205,666 & 4 \\
\multirow{-3}{*}{H2: Prolificness} & During & Number of blobs in the hackathon project repo that were created during the event. & \WOC & 1 - 3,288 & 23 \\\hline
 & pctCode & Fraction of the blobs in the project repo that are classified as ``programming''. & \WOC and \GH & 0 - 1 & 0.40 \\
 & pctMarkup & Fraction of the blobs in the project repo that are classified as ``Markup''. & \WOC and \GH & 0 - 0.96 & 0.02 \\
 & pctData & Fraction of the blobs in the project repo that are classified as ``Data''. & \WOC and \GH & 0 - 0.999 & 0.12 \\
\multirow{-4}{*}{H3: Composition} & pctProse & Fraction of the blobs in the project repo that are classified as ``Prose''. & \WOC and \GH & 0 - 0.999 & 0.03 \\ \hline
H4: Permission & LicenseType & Type of License used in hackathon project. & \WOC and \GH & \multicolumn{2}{p{3cm}}{\cellcolor[HTML]{FFFC9E}NOLICENSE :10,045 (77\%), OSSLICENSE: 1,882 (15\%), Other: 1,081 (8\%)} \\
\bottomrule
\end{tabular}%
}
%\vsapce{-15pt}
\end{table}

In addition to tracking hackathon code blob reuse (RQ2), we also aimed to study factors that can affect this phenomenon. For this purpose, we collected various characteristics of the hackathon projects, both from \DP and \WOC, and extracted the variables of interest, per the hypotheses presented in \cref{sec:rq}.

The data we collected for RQ2 was for the blobs, so, in order to find out if code blobs from a project were reused, we investigated how many blobs from a project was reused, and calculated the ratio of the number of reused code blobs and the total number of code blobs in the project. This revealed that almost 60\% of projects had none of their code blobs reused. So, we decided to pursue a binary classification problem for predicting if a project has at least one code blob reused or not instead of doing regression analysis.

For the purpose of our analysis, we excluded hackathon projects with a single member, since a hackathon project ``team'' with a single participant does not really make a lot of sense, and also the projects that were not related to any existing technology, since these likely were non-technical events. By looking for code blob reuse, we also automatically filtered out any project that had no code blobs in its repository. Moreover, projects for which we had missing values for any of the variables of interest (as shown in \cref{t:variables}) were also removed. After these filterings, we were left with 13,008 hackathon projects.

For the variables related to \textbf{H3}, the composition of the repository, we have 5 categories, 4 of which are dictated by the \GH \textit{linguist} tool: \\\textit{Code(programming), Markup, Data}, and \textit{Prose}, and a category \textit{Other} for all file types not classified by the tool. We looked at what percentage of the blobs in the projects belonged to which type. Since they all sum up to 100\%, 4 of these variables are sufficient to describe the fifth variable. In order to remove the resulting redundancy, we decided to remove the entry for type \textit{Other}, since its effect is sufficiently described by the remaining variables.

As for the license variable associated with \textbf{H4}, not all licenses used by the hackathon projects are Open Source licenses allowing the code to be shared and used by others and the projects without a license are, by default, assumed to be not licensed for reuse, i.e. under exclusive copyright. In order to capture that, we categorized the variables into three categories - projects without a license were put into the ``NOLICENSE'' category, projects with any of the standard OSS license that allow code-sharing (with some restrictions in some cases, e.g. the ``copyleft'' clause for GNU GPL) were put in the ``OSSLICENSE'' category, and projects with non-standard licenses that allow code sharing under very specific conditions or explicit copyright notices were put in the ``Other'' category.

The description of all the variables along their sources and values are presented in \cref{t:variables}.

\paragraph{Analysis Method for Identifying project characteristics that affect code blob reuse}
\label{sss:rq3-analysis}
As we noted in the hypotheses presented in \cref{sec:rq}, we are expecting some of the project characteristics to have a linear effect on hackathon code blob reuse, while some should have a more complex non-linear effect. The goal of our analysis is not to make the best predictive model that gives the optimum predictive accuracy, instead, we are trying to find out which of the predictors have a significant effect by creating an explanatory model. As noted by Shmueli~\cite{shmueli2010explain}, these two are very different tasks.

In order to achieve our goal of having linear and non-linear predictors in the same model and be able to infer the significance of each of them, we decided to use Generalized Additive Models (GAM). Specifically, we used the implementation of GAM from the \texttt{mgcv} package in \textit{R}. 

\subsection{Surveys}
\label{sec:method:survey}
In addition to the data-driven study described above, we also conducted a survey study to validate our findings and further investigate where the code blobs used in a hackathon originates from (RQ 1.c) and aspects of what happens to the code blob that was created during an event after it had ended (RQ 2). In order to answer these research questions we focused on three separate scenarios:
\begin{enumerate}
    \item The perception of individuals who had created a code blob during a hackathon that was reused after the event,
    \item The perception of individuals who had created at least a code blob during a hackathon, and none of the blobs they created in that hackathon was reused after the event, 
    \item The perception of individuals who reused a code blob that was created during a hackathon.
\end{enumerate}
These scenarios are useful because they allow us to compare the perceptions of individuals who created code that got reused with those whose code did not, and also assess the perception of individuals about code blobs they reused which was originally created during a hackathon. The survey instrument focuses on commits that contain specific blobs that fit the three aforementioned scenarios.

\subsubsection{Survey design}
We developed two separate survey instruments to cover the three scenarios. The first focuses on code blobs that was created during a hackathon thus covering the first two scenarios. The second focuses on code blobs that was originally developed during a hackathon and subsequently reused thus covering the third scenario. In the following we will outline the content of both survey instruments (Tables~\ref{tab:app:instruments:survey1} and~\ref{tab:app:instruments:survey2} in appendix~\ref{sec:app:instruments} provides an overview).

Although the same survey instrument was used for potential participants from scenarios 1 and 2, for ease of reference, we refer to them as survey 1 and survey 2, and the survey sent out to potential participants in scenario 3 as survey 3.
All of the surveys started by asking participants about whether they authored the commit we had attributed to them or not. The aim of this question was to ensure that the respective participant was knowledgeable about the code contained in the blob the survey focused on.

The second question in the surveys focused on the source of the code contained in a blob (RQ 1.c). We provided a list of options that participants could choose from which included options like ``\textit{From other GitHub repositories}'', ``\textit{From the web (e.g. on Stackoverflow or forums)}'' and ``\textit{It was generated by a tool}''. Participants could select multiple sources since it is possible that the code contained in a blob is based on different sources. Participants could also provide an alternative source using a text field. The first two surveys also included options related to the hackathon the code was created in including ``\textit{I wrote it during the hackathon}'', ``\textit{I reused code that I had written before the hackathon}'' and ``\textit{My team members wrote it during the hackathon}''.

The first two surveys continued with questions examining whether and where participants use their own code after an event and how they foster code reuse (RQ 2.d). Possible answer options included common reuse settings such as ``\textit{another hackathon}'', ``\textit{school as part of a class, project or thesis}'' or ``\textit{In another open source project in my free time}''. In addition, participants could state that they did not reuse the code after an event or mention other possible scenarios. Regarding the question of how participants fostered reuse, we included typical approaches such as ``\textit{sending it to friends or colleagues}'', ``\textit{sharing it online (e.g. Stackoverflow, Medium, Social media)}'' as well as the option to state a different approach.

We also asked participants in the first two surveys for whom they thought the code they created would be useful and whether they are aware if anyone used their code after the event (RQ 2.d). Possible answer options regarding their perception of the usefulness of the code included ``\textit{No one}'', ``\textit{For our hackathon project or team}'' and ``\textit{For many people outside of the hackathon project or team}''. Possible options related to their awareness about who might have reused their code included ``\textit{my hackathon team members}'', ``\textit{my friends or colleagues}'' and ``\textit{someone else used it in an open-source project}''. For both questions, participants could again state other possible options using a text field.

In addition to asking about a specific blob, we also included questions regarding the perception of the participant about the hackathon project during which they developed to code in the blob in the first surveys. We specifically focused on their perception of the usefulness of the project and their intentions to continue working on it (RQ 2.a and RQ 2.b). For both aspects, we relied on established scales~\cite{reinig2003toward,bhattacherjee2001understanding} which and have been successfully applied in research related to hackathons in the past~\cite{filippova2017diversity,nolte2020support}.

In addition to the code that was developed participants might also have reused the idea their project was based on in a different context (RQ 2.f). The first surveys thus also included a simple \textit{yes/no/not sure} question related to this aspect.

Regarding the third survey, we also included a question that focused on how easy they found the using the code contained in the blob we identified (RQ 2.a). For this, we adapted an established scale by Davis~\cite{davis1989perceived}.

Finally, all surveys also included common demographic questions, including the age of the participant, their gender, their perception about their minority status, and their experience contributing to open-source projects. We utilized this simple way of assessing perceived minority since minority perception of an individual can be based on many different aspects like race, gender, expertise, background, and others. Table \ref{tab:demographics} shows summary for the participants demographics.

\begin{table}[!h]
\caption{Surveys demographics summary}
\label{tab:demographics}
\resizebox{\textwidth}{!}{%

\begin{tabular}{cccccc}
\hline
 &  & \textbf{\begin{tabular}[c]{@{}c@{}}Survey 1\\ Percentage \\ \%\end{tabular}} & \textbf{\begin{tabular}[c]{@{}c@{}}Survey 2\\ Percentage \\ \%\end{tabular}} & \textbf{\begin{tabular}[c]{@{}c@{}}Survey 3\\ Percentage \\ \%\end{tabular}} & \textbf{\begin{tabular}[c]{@{}c@{}}Combined \\ Percentage \\ \%\end{tabular}} \\ \hline
\multirow{4}{*}{\textbf{Gender}} & Male & 83.5 & 79.3 & 85.3 & 82.8 \\ \cline{2-6} 
 & Female & 11.9 & 19.0 & 10.3 & 13.5 \\ \cline{2-6} 
 & Prefer not to say & 4 & 1.7 & 4.3 & 3.4 \\ \cline{2-6} 
 & Non-binary & 0.6 & 0 & 0 & 0.2 \\ \hline
\multirow{6}{*}{\textbf{Age Group}} & 18 to 24 & 49.7 & 61.0 & 41.4 & 50.6 \\ \cline{2-6} 
 & 25 to 34 & 40.7 & 35.6 & 42.2 & 39.7 \\ \cline{2-6} 
 & 35 to 44 & 6.8 & 2.5 & 6.9 & 5.6 \\ \cline{2-6} 
 & 45 to 54 & 0 & 0 & 5.2 & 1.5 \\ \cline{2-6} 
 & 55 to 64 & 0 & 0.8 & 0 & 0.2 \\ \cline{2-6} 
 & Prefer not to say & 2.8 & 0 & 4.3 & 2.4 \\ \hline
\multirow{5}{*}{\textbf{Experience}} & 1 & 18.4 & 25.0 & 14.1 & 19.0 \\ \cline{2-6} 
 & 2 & 23.4 & 22.7 & 21.7 & 22.7 \\ \cline{2-6} 
 & 3 & 12.8 & 17.0 & 20.7 & 16.2 \\ \cline{2-6} 
 & 4 & 14.2 & 11.4 & 15.2 & 13.7 \\ \cline{2-6} 
 & 5+ & 31.2 & 23.9 & 28.3 & 28.3 \\ \hline
\multirow{3}{*}{\textbf{Minority}} & Yes & 30.2 & 41.1 & 25.5 & 32.0 \\ \cline{2-6} 
 & No & 64.5 & 56.2 & 70.9 & 64.0 \\ \cline{2-6} 
 & Rather not say & 5.2 & 2.7 & 3.6 & 4.1 \\ \hline
\end{tabular}
}
\end{table}

\subsubsection{Procedure}\label{ss:survey-proc}
We selected 5,000 individuals to contact for each of the previously discussed scenarios. For each of them, we followed the following procedure - We first selected blobs appropriate for the respective scenario and that contained more than 15 lines of code. We then identified the commits that they were contained in and removed all blobs which could not be reached in \GH. We proceeded to remove duplicates, sorted the commits by date, and selected the 5,000 most recent ones per scenario. Finally, we ensured that no individual would receive an invitation for more than one survey.

Before sending the surveys we piloted them with 3 individuals to ensure that they are understandable and to remove potential misspellings and other mistakes. We then sent individual invitations to the selected participants. Out of the 15,000 invitations 3,254 could not be delivered (1,036 for survey 1, 1,083 for survey 2, and 1,135 for survey 3). We received a total of 1,479 responses (558 for survey 1, 335 for survey 2, and 586 for survey 3) which amounts to a total response rate of 9.86\%. We subsequently removed incomplete responses and responses that lacked appropriateness (e.g. if someone stated that they did not contribute the commit we had attributed to them). We arrived at a total of 416 complete and useful responses (178 for survey 1, 120 for survey 2, and 118 for survey 3) thus having an effective response rate of 2.77\%.

In addition to this cleaning process, we also assessed the internal consistency of the three scales we utilized using Cronbach's alpha~\cite{tavakol2011making}. The respective values were at 0.865 for perceived usefulness, 0.841 for continuation intentions, and 0.934 for perceived ease of use which makes all scales suitable for further analysis.

The analysis was mainly based on comparing findings from the surveys related to code blobs that were reused after a hackathon (survey 1) and code blobs that were not (survey 2). We particularly compared code origins (RQ 1.c), reuse behavior by code creators (RQ 2.c), code sharing behavior (RQ 2.d), awareness about code reuse, and perceptions about the usefulness of the code (RQ 2.e). In addition, we also compared satisfaction perceptions and continuation intentions on a project level (RQ 2.a and RQ 2.b) as well as reuse of the ideas that arose from the hackathon projects (RQ 2.f). All comparisons were done on a descriptive level.
Finally, contributing to RQ 2.a, we also analyzed the perception of participants related to how easy they found using the code based on survey 3.

% \tapd{please check if this fits and adapt accordingly}

\section{Results}
\label{sec:results}

Here we will discuss our findings in relation to our research questions.

\subsection{Origins of hackathon code blobs (RQ1)}
\label{sec:results:rq1}

\begin{figure}[!t]
\centering
%\vsapce{-10pt}
\includegraphics[width=\linewidth]{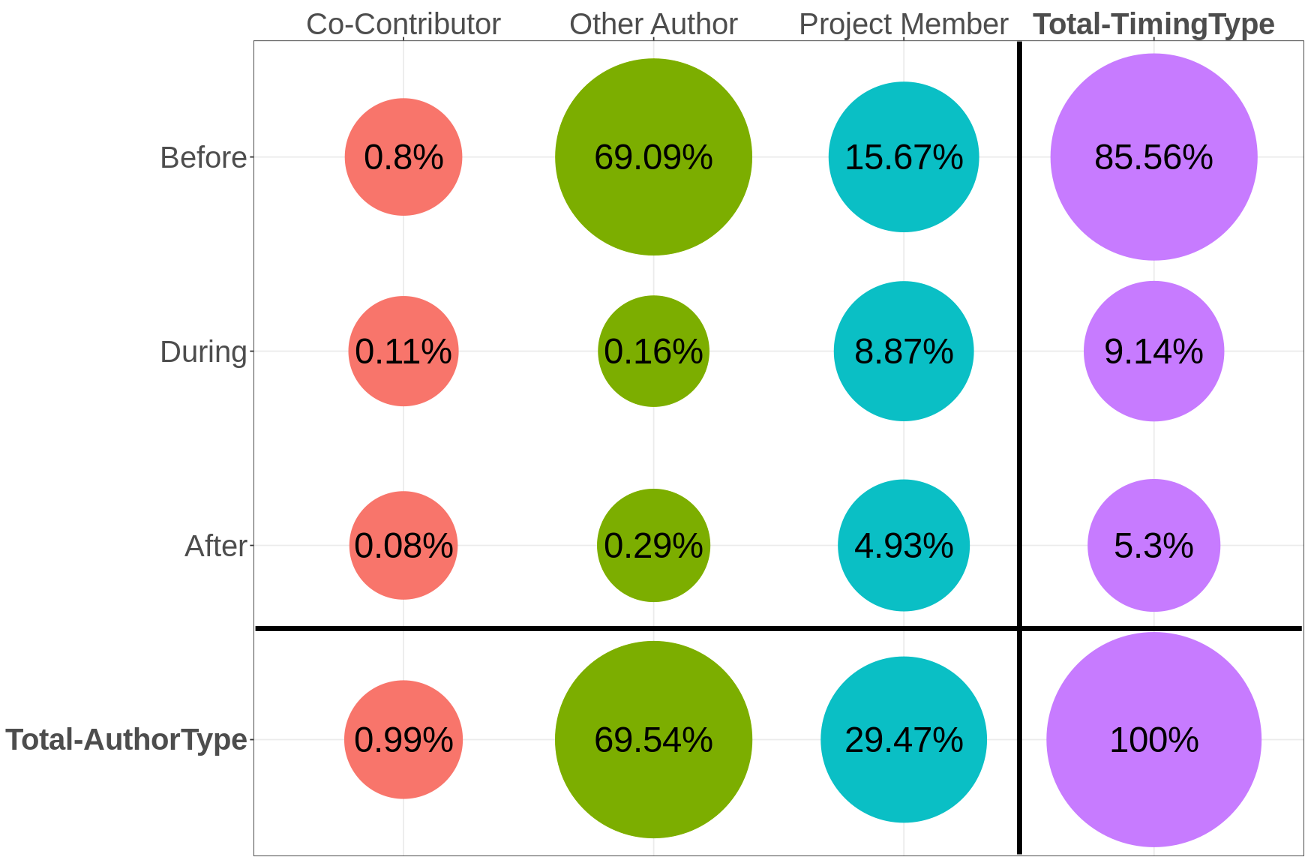}%
%\vsapce{-10pt}
\caption{\textbf{Plot of Who created how much of the Hackathon Code and When \\(RQ 1.a and RQ 1.b)}}
\label{fig:rq1}
%\vsapce{-15pt}
\end{figure}

As mentioned in \cref{sec:rq}, our first research question focuses on several aspects related to the \textit{origin} of the hackathon code. We draw insights from both the data-driven study on hackathon projects as well as the surveys to address this research question.

\subsubsection{When are code blobs used in hackathon projects created (RQ 1.a) and who are the original authors (RQ 1.b) of the code blob?}

In terms of ``when'', we examined if the first creation of the code blob under consideration was \textit{Before, During}, or \textit{After} the corresponding hackathon event. In terms of ``who'', we checked if the first creator of the code blob was one of the members of the hackathon project (\textit{Project Member}) or someone who was a contributor to a project in which one of the members of the hackathon project contributed to as well (\textit{Co-Contributor}), or someone else (\textit{Other Author}).

The result of the analysis is presented in \cref{fig:rq1}, showing that, overall, 85.56\% of the code blobs used in the hackathon projects was created before an event. Most of these reused blobs were part of a framework/library/package used in the hackathon project, which aligns with the findings of~\cite{mockus2007large}. Around 9.14\% of the blobs were created during events, since participants need to be efficient during an event owing to the time limit~\cite{nolte2018you} which fosters reuse as previously discussed in the context of OSS~\cite{haefliger2008code}. We also found that 5.3\% of the blobs were created after an event, suggesting that most teams do not add a lot of new content to their hackathon project repositories after the event. This finding is in line with prior work on hackathon project continuation~\cite{nolte2020what}. 

\hypobox{\textbf{RQ1.a}: 85.56\% of the code (in terms of the no. of blobs) in the hackathon project repositories is created before the hackathons, with around 9.14\% of the code blobs being created during the events (which is significant considering the limited duration of the hackathons).} 

Figure \ref{fig:rq1} shows that, overall, the original creators of most of the code blobs (69.54\%) in the hackathon project repositories are someone who is not a part of the team. They are mostly the original creators of some project/package/framework used by the hackathon team. Around one-third (29.47\%) of the code blobs were created by the project members, and the reuse of code blobs from co-contributors in other projects is very limited (0.99\%). This aspect has not been extensively studied in the context of work on hackathons yet.

\hypobox{\textbf{RQ1.b}: The members of the hackathon teams created around 29.47\% of the code blobs, while 69.54\% of the code blobs are created by developers outside the team (mostly authors of some project/package/framework used by the team).}

\subsubsection{Where did the code come from? (RQ 1.c)}

\begin{table}[!h]
\caption{Code Origin Response from surveys of hackathon participants whose code was reused (Survey 1) and whose code wasn't reused (Survey 2)}
\label{tab:CodeOriginSurveys}
\resizebox{\textwidth}{!}{%

\begin{tabular}{@{}cccccc@{}}
\toprule
\begin{tabular}[c]{@{}c@{}}I wrote it \\ during \\ the hackathon\end{tabular} & \begin{tabular}[c]{@{}c@{}}Team members \\ wrote it \\ during \\ the hackathon\end{tabular} & \begin{tabular}[c]{@{}c@{}}From \\ the web\end{tabular} & Others & \begin{tabular}[c]{@{}c@{}}Count\\ Survey1\end{tabular} & \begin{tabular}[c]{@{}c@{}}Count\\ Survey2\end{tabular} \\ \midrule
\xmark&\xmark&\xmark&\xmark&0&0\\
\xmark&\xmark&\xmark&\cmark&22&8\\
\xmark&\xmark&\cmark&\xmark&6&0\\
\xmark&\xmark&\cmark&\cmark&0&0\\
\xmark&\cmark&\xmark&\xmark&7&6\\
\xmark&\cmark&\xmark&\cmark&0&0\\
\xmark&\cmark&\cmark&\xmark&2&0\\
\xmark&\cmark&\cmark&\cmark&2&1\\
\cmark&\xmark&\xmark&\xmark&76&60\\
\cmark&\xmark&\xmark&\cmark&6&5\\
\cmark&\xmark&\cmark&\xmark&20&12\\
\cmark&\xmark&\cmark&\cmark&4&7\\
\cmark&\cmark&\xmark&\xmark&15&11\\
\cmark&\cmark&\xmark&\cmark&3&3\\
\cmark&\cmark&\cmark&\xmark&6&4\\
\cmark&\cmark&\cmark&\cmark&9&2\\
\bottomrule
\end{tabular}
}
\end{table}

Our research question RQ 1.c focuses on the origin of code contained in the blobs under consideration in order to understand how the hackathon teams generated the code written during the hackathon. This question can only be answered by directly asking the participants, thus one of the survey questions was targeting that topic by asking about the source of the code committed during the hackathon in the hackathon repository.
By analyzing the 178 and 120 responses of that question from survey scenarios 1 and 2 respectively (since they were the hackathon participants who created the code), we found that most participants chose that they wrote the code during the hackathon, team members wrote the code during the hackathon, and/or they got the code from the web. Other options, like \textit{I reused code that I had written before the hackathon, My team members reused code they had written before the hackathon, It was generated by a tool, From other GitHub repositories} were chosen rarely. Table \ref{tab:CodeOriginSurveys} contains a summary of code origin responses from surveys of hackathon participants whose code was reused (Survey 1) and whose code wasn't reused (Survey 2). Since only 3 options were chosen most frequently, we combined all other responses in one group called ``Others'' which include all non commonly chosen options plus the free text option where participants can write something not listed in the options.

The most frequently chosen option was that they wrote the code themselves (76 (42.7\%) and 60 (50\%) exclusively for Survey 1 and 2 respectively). A number of participants combined 2 options and choose that they wrote the code during the hackathon along with getting the code from the web (20 and 12 for Survey 1 and 2 respectively), which is a common practice during hackathons. Similarly, some participants choose that they write the code during the hackathon along with team members wrote it during the hackathon which indicates teamwork and collaboration during such hackathon events. Some participants provided extra information in the free text option, specifically, 9 and 5 participants utilized the free text field in survey 1 and survey 2 respectively. Participants indicated that some code files were generated by a bot (Microsoft Bot Framework Workshop), setup files for a Flask project, a dataset for fake news competition, code file provided by the hackathon organizers, or they used/adapted examples they discovered through Google search.

\hypobox{\textbf{RQ1.c}: Most (45.6\% overall) of the hackathon participants indicated (through surveys) that they wrote the code contained in the blobs under consideration themselves during the hackathon and a number of them also indicated they found the code from the web or wrote it together with another hackathon team member.}

\subsubsection{Programming languages distribution for hackathon code with different origin (RQ 1.d)}

\begin{figure}[!ht]
\centering
%\vsapce{-10pt}
\includegraphics[width=\linewidth]{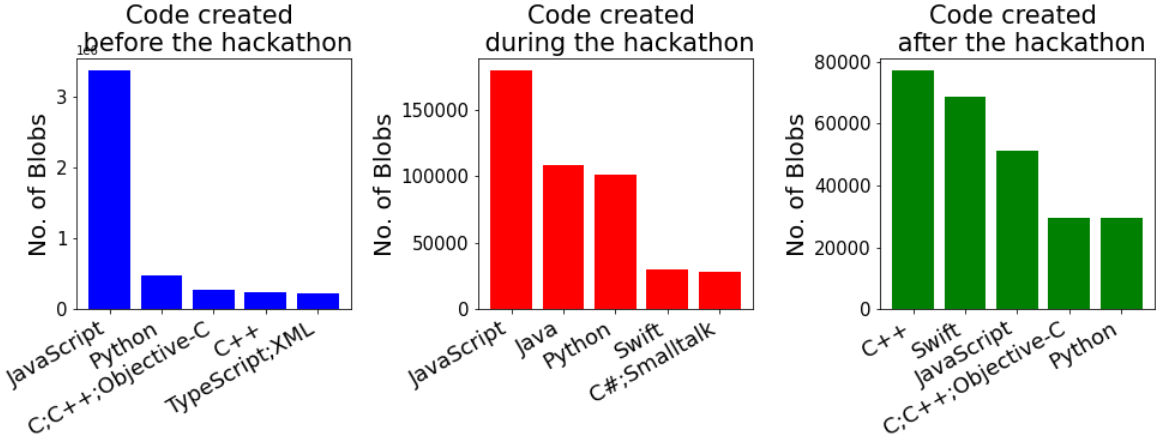}%
%\vsapce{-10pt}
\caption{\textbf{Top 5 languages for blobs created before, during, and after hackathons}}
\label{fig:rq1-time}
%\vsapce{-15pt}
\end{figure}

\begin{figure}[!ht]
\centering
% %\vsapce{-10pt}
\includegraphics[width=\linewidth]{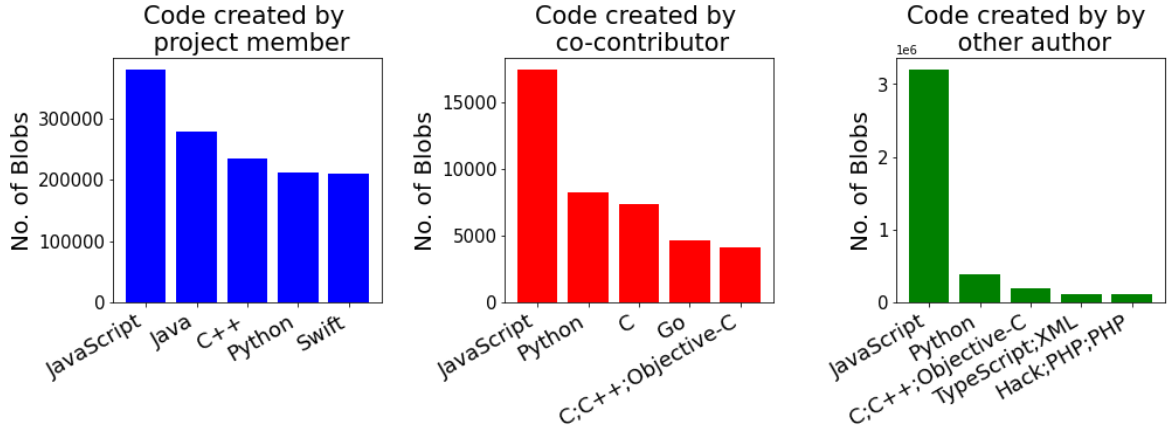}%
%\vsapce{-10pt}
\caption{\textbf{Top 5 languages for blobs created by project members, co-contributors, and others}}
\label{fig:rq1-author}
%\vsapce{-15pt}
\end{figure}

Looking at top languages for the blobs created at different times (\cref{fig:rq1-time}), identified by the \GH \textit{linguist} tool, we found that most of the code reused by hackathon projects (created before) is  JavaScript, and other top languages together indicate that most of the reused code by hackathon projects are related to web development frameworks. JavaScript is also the most common language worked on during the hackathons we studied, followed by Java, Python, Swift, and C\#/Smalltalk, indicating that most hackathon projects work on developing web/mobile apps. C++ was the most common language for code developed after the event, followed by Swift and JavaScript, showing a slight shift in the type of work done after the event, favoring Machine Learning applications.

Looking at the top languages for the code created by different authors, as shown in \cref{fig:rq1-author}, we can see that, once again, most of the code created by developers not part of the hackathon team is JavaScript, which is similar to the code created before the event (\cref{fig:rq1-time}). This is not surprising, since they have a great deal of overlap (\cref{fig:rq1}). Most of the code created by project members indicates a leaning towards web/mobile app development, and most of the C++ and Python code was found to be related to Machine Learning frameworks.

We noticed that JavaScript is, by far, the most widely used language in terms of blob count. Most of the JavaScript blobs were, in fact, related to some framework and were developed by people not part of the hackathon team. One possible reason for that could be that developing web applications or designing front-end UIs is a common hackathon activity and hackathon participants often reuse standard frameworks for developing their UIs.

\hypobox{\textbf{RQ1.d}: JavaScript is the most commonly used language for code created before or during the hackathon, followed by Java and Python and the same trend is seen when looking at who created the code. Code written in the C-family of languages is also quite common and is the most commonly used language after the hackathons. }

\subsubsection{Size distribution of hackathon code blobs (RQ 1.e)}

\begin{figure}
\centering
\begin{subfigure}{.5\textwidth}
  \centering
  \includegraphics[width=\textwidth]{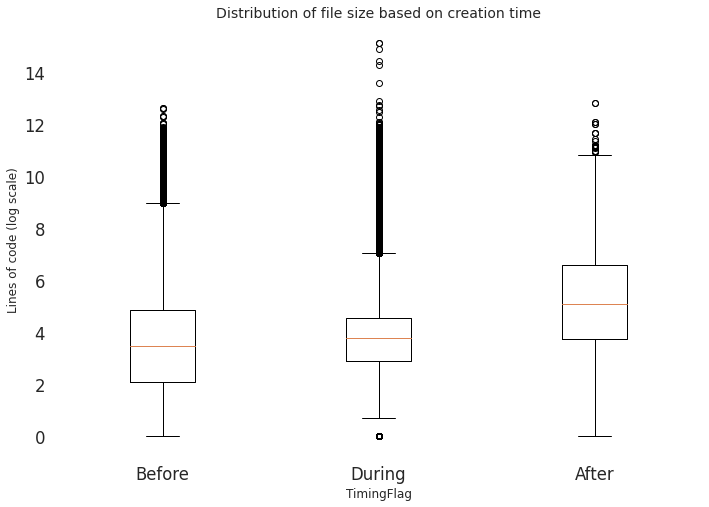}
  \caption{Different Creation Time}
  \label{fig:loc-timingflag}
\end{subfigure}%
\begin{subfigure}{.5\textwidth}
  \centering
  \includegraphics[width=\textwidth]{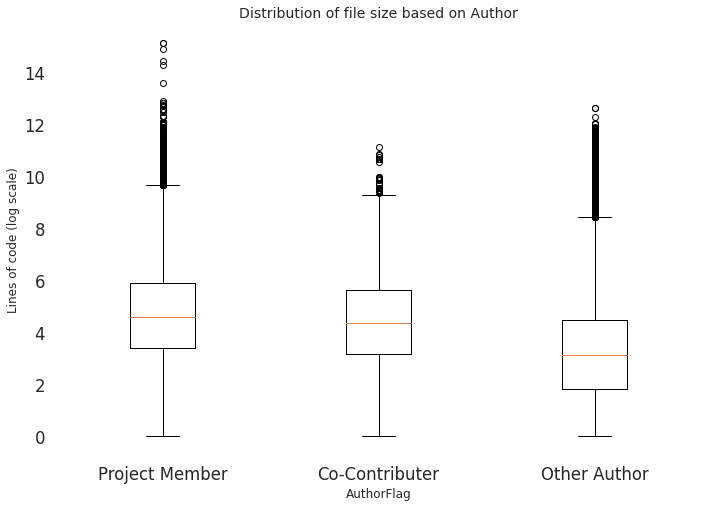}
  \caption{Different Author}
  \label{fig:loc-authorflag}
\end{subfigure}%
\caption{\textbf{Distribution of Lines of Code for code blobs created at different times (a) and by different authors (b)}}
\label{fig:loc}
\end{figure}

\begin{table}[!h]
\caption{\textbf{Descriptive statistics of code sizes (LOC) for blobs created at different times and by different authors}}
\label{t:sigdif-RQ1}
\resizebox{\textwidth}{!}{%
\begin{tabular}{@{}lccccc@{}}
\toprule
Type & Origin & Minimum LOC & Mean LOC & Median LOC & Maximum LOC \\ \midrule
\multirow{3}{*}{Timing} & Before & 0 & 241.59 & 32 & 297,827 \\ \cmidrule(l){2-6} 
 & During & 0 & 208.67 & 43 & 3,665,363 \\ \cmidrule(l){2-6} 
 & After & 0 & 772.62 & 159 & 359,982 \\ \midrule
\multirow{3}{*}{Author} & Project Member & 0 & 565.97 & 95 & 3,665,363 \\ \cmidrule(l){2-6} 
 & Co-Contributer & 0 & 362.35 & 75 & 67,233 \\ \cmidrule(l){2-6} 
 & Other Author & 0 & 142.34 & 22 & 297,827 \\ \bottomrule
\end{tabular}%
}
\end{table}

As mentioned in section~\ref{sss:loc}, we analyzed the sizes of the hackathon code blobs by measuring the LOC (lines of code). The distributions of code sizes are shown in \cref{fig:loc}, with \cref{fig:loc-timingflag} showing the code size distributions for blobs created before, during, and after a hackathon event and \cref{fig:loc-authorflag} showing the distributions for blobs created by project members, co-contributors, and others.

The first thing we checked was if the distributions are significantly different from one another. Therefore, we conducted pairwise Mann-Whitney U tests on blobs created before, during, and after the hackathon and also on code created by project members, co-contributors, and others. The results clearly indicated that the sizes of blobs created at different times and by different authors are indeed significantly different, with the p-Values for each test being $<1e-100$.

The descriptive statistics for sizes of blobs created at different times and by different authors is presented in \cref{t:sigdif-RQ1}. It goes to show that project members tend to create more lines of code on average than the rest. Although we saw a number of large outliers in the code created during the hackathons, on average, the sizes of code blobs added after the hackathons also seem to be larger. It is worth noting, however, that we only looked at the lines of code and not, for example, the complexity of the code. For example, it could very well be the case that the code created by project members is larger because they are not optimized very well due to the time constraints of the hackathons. 

One curious oddity we noticed is that the minimum LOC for blobs in every category is 0 and we further investigated those blobs. Upon examination, we found 1,150 unique blobs among the ones under study that were identified as code blobs but had 0 lines of code. However, almost all of these blobs were found to contain some commented code which led us to believe that these blobs might have been used sometime during development, e.g. for testing a feature or debugging, and were later commented out because e.g. the team didn't implement the feature or finished the debugging task.

Overall, we found the total size of code blobs created before, during, and after the hackathon to be $1,125,735,653$, $110,921,111$, and $203,786,274$ respectively. Therefore, considering the total lines of code, 78\% of code used in the hackathon projects were created before the hackathons, 8\% during the hackathons, and 14\% after the hackathons. Similarly, a total of $871,575,852$ (60.5\%) lines of code were created by the hackathon project members, $19,147,025$ (1.3\%) lines of code were created by the co-contributors, and $549,720,161$ (38.2\%) lines of code were created by other developers.

\hypobox{\textbf{RQ1.e}: Around 8\% of code is created during the hackathons in terms of the total lines of code, but around 60.5\% of the total lines of code are created by the project members. The distributions of code sizes for blobs created at different times are significantly different from one another and so are the sizes of blobs created by different authors. The sizes of blobs created after the hackathons tend to be larger on average, and project members also seem to create larger blobs than the others.}

\subsubsection{Identifying template code (RQ 1.f)}

\begin{figure}[!ht]
\centering
%\vsapce{-10pt}
\includegraphics[width=\linewidth]{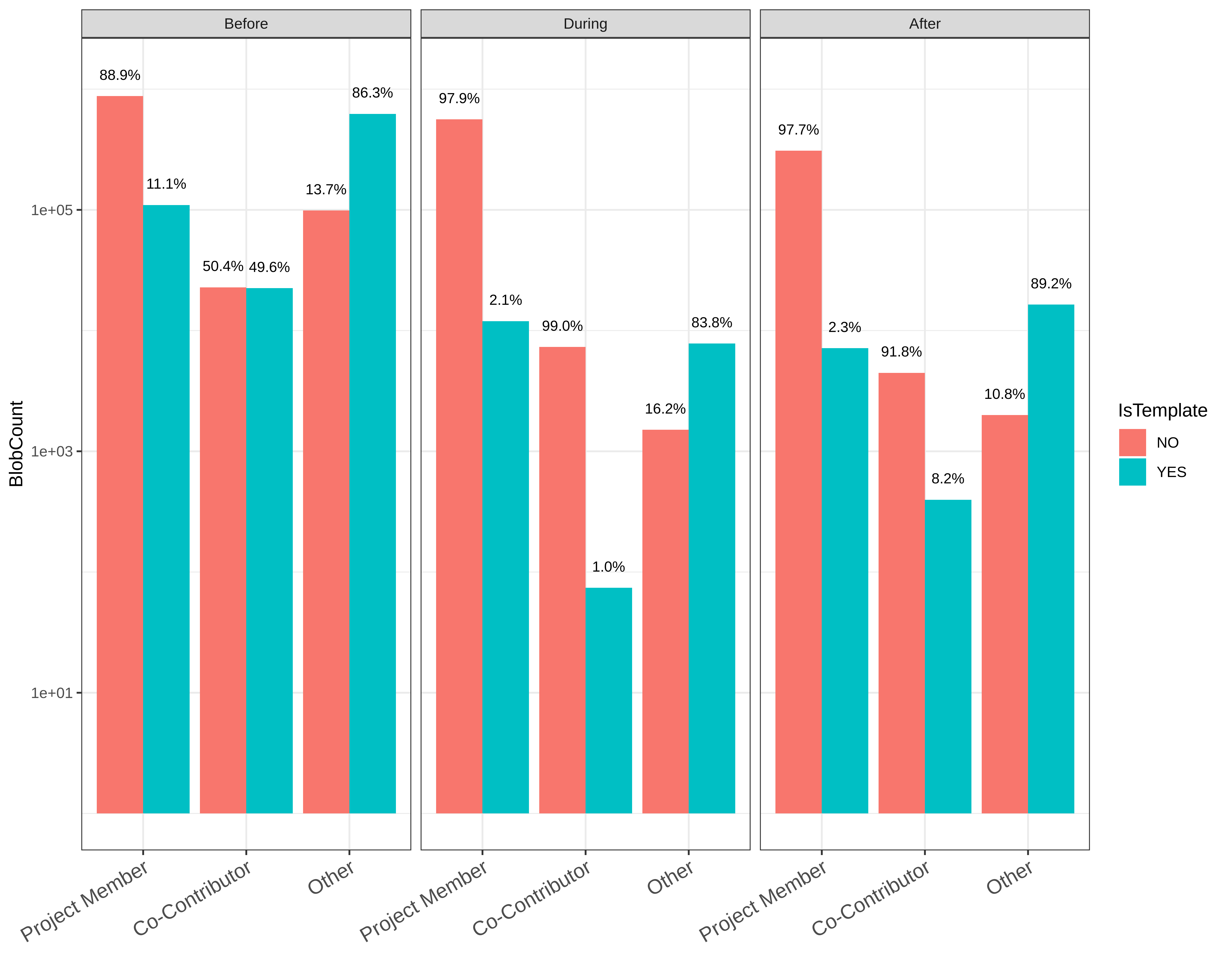}%
%\vsapce{-10pt}
\caption{\textbf{Plot of the distribution of Template Code depending on Who created the code and When}}
\label{fig:rq1-template}
%\vsapce{-15pt}
\end{figure}

Using the method described in section~\ref{sss:template} we analyzed 2,684,870 code blobs used in hackathon projects to check if they were template code blobs or not. Our heuristics rely on a blob-to-filename (\textit{b2f}) map in order to determine that and the mapping was missing for a number of blobs leading us to exclude them from our analysis. Overall, we found that 800,294 (29.8\%) blobs were identified as template code by our method.

The distribution of template code used in the hackathon projects depending on who created the code blob (a project member, a co-contributor, or other developers) and when was the code blob first added to the hackathon repository (before, during, or after the hackathon event) is shown in \cref{fig:rq1-template}. 
The three subplots refer to the timing and the X-axis values in each subplot refer to the author of the code.
The percentage numbers show how much of the code is template code (or not) for each author-timing combination. 

We notice that most of the code (in terms of blob count) hackathon project members write before, during, and after the event are non-template code and, conversely, most of the code created by the ``other'' developers used in the hackathon projects is actually template code. Overall, most of the code blobs added to the hackathon projects before the events is template code while the code blobs added during or after the event is mostly non-template code. This shift is even more drastic if we look at the code created by the co-contributors - while the code created by them is almost evenly divided between template and non-template code before the event, most of the code created by them that is added to the hackathon projects during or after the hackathon event are non-template code. This goes to show that the hackathon participants tend to use templates created by other developers in their projects quite often but they themselves create a very small amount of template code, which aligns with our experience about how hackathon events unfold. It is also interesting that hackathon teams do not reuse a lot of non-template code - one might assume that during the hackathons teams might try to use existing code written by other developers working on a similar problem but that doesn't seem to be true in most cases.

\hypobox{\textbf{RQ1.f}: Most of the code (blobs) created by project members is non-template code while most of the code (blobs) created by developers outside of the team tends to be template code. This trend is consistent for code blobs added to the projects before, during, or after the corresponding hackathon events.}

% \tapd{Summary finding of RQ1}
% Therefore, to give a more complete answer to RQ1:
% \hypobox{\textbf{Origin of the Hackathon Code (RQ1)}: Hackathon projects often reuse code in terms of some package/framework. Teams also tend to reuse their own code. Most of the code created during or after the event is created by the hackathon team members.}

\subsection{Hackathon code blob reuse (RQ2)}
\label{sec:results:rq2}

%\begin{figure}[!htb]
%\centering
%%\vsapce{-10pt}
%\includegraphics[width=\linewidth]{RQ2-max.png}%
% %\vsapce{-10pt}
%\caption{Plot depicting Hackathon code reuse in projects of different categories %when only the maximum size is considered}
%\label{fig:rq2-max}
% %\vsapce{-10pt}
%\end{figure}

As discussed in \cref{sec:rq} and section \ref{sss:rq2}, our goal while looking for hackathon code blob reuse is to understand the reuse of code blobs that were first created during a hackathon event.

\subsubsection{Comparing the characteristics of code blobs that were reused (RQ 2.a) and code blobs that weren't (RQ 2.b)}

By following the procedure outlined in Section \ref{sss:rq2}, we found that 167,781 (28.8\%) of the 581,579 hackathon code blobs got reused in other projects. 

%\ai{Add size distribution of reused and not reused code blocks as a pair of boxplots}
% \tapd{Done, please check}

\begin{figure}[!h]
\centering
%\vsapce{-10pt}
\includegraphics[width=0.8\linewidth]{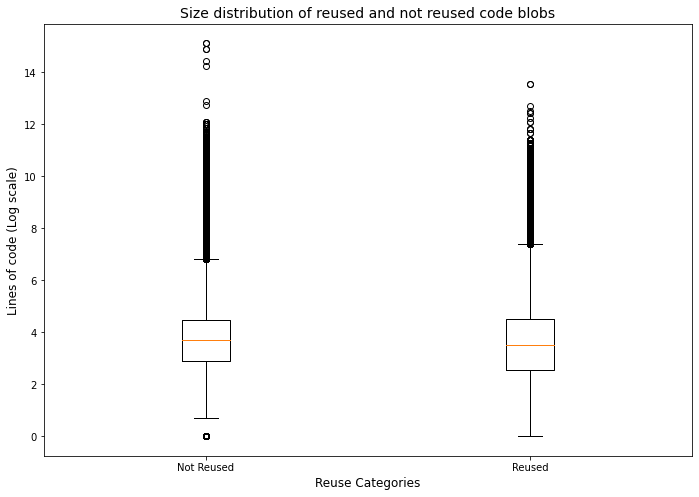}%
%\vsapce{-10pt}
\caption{\textbf{Size distribution of reused and not reused code blobs}}
\label{fig:sizeDist-Reused}
%\vsapce{-15pt}
\end{figure}

A quick analysis into whether the shared blob is a template code blob or not revealed that, in fact, 159,286(95.5\%) of the blobs that were reused later on were not ``template'' code. In contrast, around 29.8\% of the blobs used in hackathon projects were template code. A reason for this apparent discrepancy could be that hackathon projects do not typically create template code. As also shown in \cref{fig:rq1-template}, only around 2.1\% of the code blobs we consider for RQ2 are template code. Given that figure, around 4.5\% of the blobs among the ones reused does seem to indicate that template blobs have a higher propensity of being reused. Moreover, only 7,842 (1.9\%) blobs among the 413,798 that were not reused later on were found to be template code blobs, which seems to further support the assertion.

\begin{table}[ht]
\centering
\caption{Logistic Regression model for statistically explaining code blob reuse by whether a blob is template or not}\label{t:reuse-template}
\resizebox{\textwidth}{!}{%

\begin{tabular}{rrrr}
  \hline
 & Estimate & Std. Error & p-Value \\ 
  \hline
(Intercept) & -0.9355 & 0.0030  & $<2e-16$ \\ 
  Template-\textbf{YES} & 1.0155 & 0.0159  & $<2e-16$ \\ 
   \hline
\end{tabular}
}
\end{table}

We decided to test this assertion by using a logistic regression model where we used if a code blob was reused or not as a binary response variable and if a code blob is template code or not as a binary predictor variable. The result of which, as shown in \cref{t:reuse-template}, verifies that that template code has a higher probability of being reused. 

\begin{table}[!ht]
\caption{Descriptive Statistics for Code sizes (LOC) for Hackathon blobs that were later reused and blobs that were not reused}
\label{t:rq2-loc}
\resizebox{\textwidth}{!}{%
\begin{tabular}{@{}cccll@{}}
\toprule
Type & Minimum LOC & Mean LOC & Median LOC & Maximum LOC \\ \midrule
Reused & 0 & 185.92 & 33 & 771,207 \\ \hline
Not-reused & 0 & 187.55 & 40 & 3,665,363\\\bottomrule
\end{tabular}%
}
\end{table}

Considering the total size of the blobs, the reused blobs contained a total of $31,662,234$ lines of code, which is around 29\% of the total $108,975,835$ lines of code contained in the 581,579 blobs under consideration. The size distribution for the relevant blobs that were later reused and the ones which were not reused is shown in \cref{fig:sizeDist-Reused} and the descriptive statistics for them is available in \cref{t:rq2-loc}. The blobs that were not reused were found to be, on average, larger than the ones that were reused. A Mann-Whitney U test revealed that the distribution of code sizes of blobs that were reused and blobs that were not is also significantly different (p-Value $<1e-100$). 

\begin{figure}[!h]
\centering
%\vsapce{-10pt}
\includegraphics[width=0.8\linewidth]{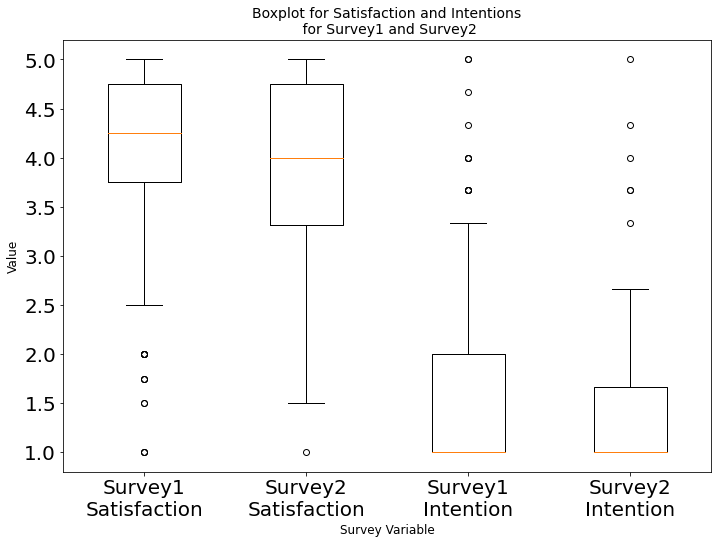}%
%\vsapce{-10pt}
\caption{\textbf{Boxplot for Satisfactions and Intentions scale questions}}
\label{fig:satInt}
%\vsapce{-15pt}
\end{figure}

Another aspect we wanted to understand is if the hackathon participants' satisfaction with the hackathon event and intention of continuing the project in the future is different depending on whether the code got reused or not. The reason we investigated these two aspects is that we thought this might have an effect on the quality of code they developed and might, consequently, affect the probability of the code being reused later on.
These aspects were investigated through the surveys we sent out. We had the participants rate five questions related to their satisfaction with the hackathon event and four questions related to their future intentions about continuing the project work on a 5-point Likert scale (full list of questions are available at \cref{tab:app:instruments:survey1} in Appendix). As discussed in \cref{ss:survey-proc}, we calculated the Cronbach’s alpha for the responses and found them to be consistent. Finally, we converted their Likert scale-based responses to numerical values and calculated the mean of responses for each participant for the questions related to satisfaction and the questions related to intention for continuation. 

The resultant distribution of responses from the participants of surveys 1 and 2 is shown in \cref{fig:satInt}, with the values in the Y-axis indicating the participants' satisfaction with the hackathon (5: Very Satisfied, 1: Not at all Satisfied) and if they planned on continuing the project (5: Very likely to continue, 1: Very unlikely to continue). We noticed from \cref{fig:satInt} that while most participants were satisfied with the hackathon events they were also unlikely to continue working on the projects. Although it looked like participants from survey 1 were, overall, more satisfied with the event and were more likely to continue working on the project compared to the participants of survey 2, the difference was found to be not statistically significant using a Mann-Whitney U test.

\begin{figure}[!h]
\centering
%\vsapce{-10pt}
\includegraphics[width=\linewidth]{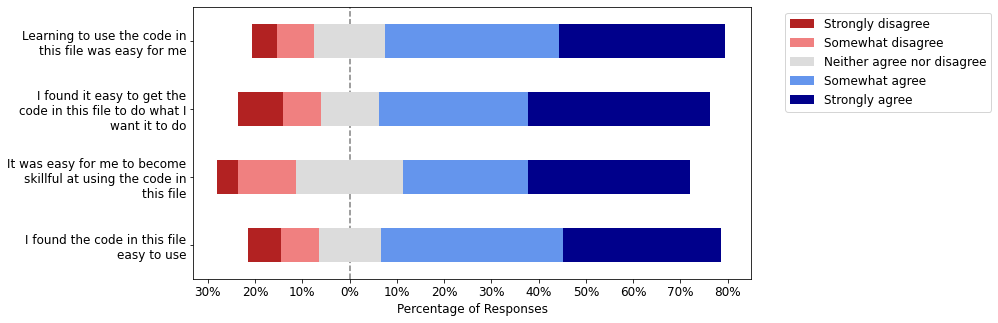}%
%\vsapce{-10pt}
\caption{\textbf{Ease of Use of hackathon code that was reused (N=118)}}
\label{fig:Sur3-EOU}
%\vsapce{-15pt}
\end{figure}

Finally, we also wanted to get a sense of how easy it is to reuse these hackathon code blobs. This can only be done credibly if we ask someone who was not part of the original team and later reused the code. For obvious reasons, we can't test this for code that wasn't reused later on. We had a series of questions in the survey that we sent out to developers who reused some hackathon code asking them to rate its ease-of-use on a 5-point Likert scale. The result, as shown in \cref{fig:Sur3-EOU}, shows that most of the participants rated the code to be easy to use across different dimensions.

\hypobox{\textbf{RQ2.a and 2.b}: Around 28.8\% and 29\% of the hackathon code got reused in other projects in terms of the number of blobs and the total line of code respectively. The characteristics of the blobs that were reused in terms of size (LOC) and whether the code (blob) is template code or not were found to be significantly different from those that were not reused, with template code having a higher chance of being reused than non-template code. By analyzing the responses from the participants of survey 1, whose code was reused, and participants of survey 2, whose code wasn't, we found no significant difference between their satisfaction with the hackathon event or future intentions of continuing to work on the project.  Participants of survey 3, who reused some hackathon code, indicated the code was easy-to-use in most cases.}

\subsubsection{Projects that reuse hackathon code (RQ 2.c)}
We further classified the projects that reused these code blobs into \textit{Small, Medium}, and \textit{Large}, as discussed in Section \ref{sss:rq2}. To recap, 89.2\% of the projects that reused the hackathon code blobs were classified as \textit{Small}, 8.5\% were \textit{Medium}, and 2.3\% were classified as \textit{Large} projects. By investigating the blobs reused by these projects we found that, unsurprisingly, there are a number of instances where a blob was reused in projects of different categories.  However, such cases were found to be quite rare, in fact, only 8.85\% of reused blobs got reused in more than one project. By looking at the size of the projects a blob was reused in, we found that over half (57.73\%) of the blobs are only reused in other \textit{Small} projects, around one-third (32.85\%) are reused in \textit{Medium} projects, and less than a tenth (9.42\%) are reused in \textit{Large} projects. 

\begin{figure}[!t]
\centering
%\vsapce{-10pt}
\includegraphics[width=\linewidth]{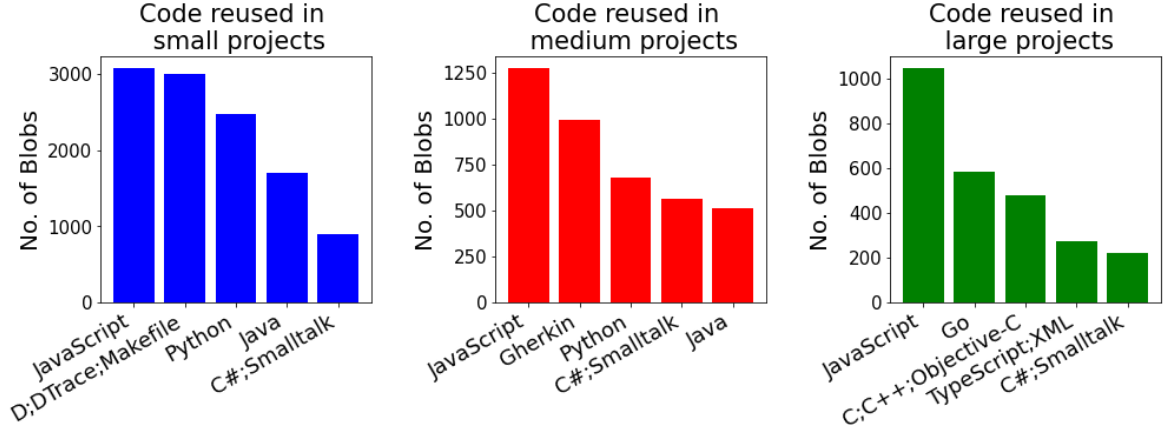}%
%\vsapce{-10pt}
\caption{\textbf{Top 5 Languages for the reused code blobs in different projects}}
\label{fig:rq2-lang}
%\vsapce{-15pt}
\end{figure}

The top-5 languages for the blobs reused by various projects are shown in \cref{fig:rq2-lang}. As we can see JavaScript still remains the most common, and Python, C/C++, C\#/Smalltalk, Java were among the top ones as well. While most reused blobs are related to web/mobile apps/frameworks, we also found the relatively uncommon Gherkin being the second most common language for \textit{Medium}  projects, and the \textit{Small} projects reused a lot of blobs related to D/DTrace/Makefile. 

\begin{figure}[!t]
\centering
%\vsapce{-10pt}
\includegraphics[width=\linewidth]{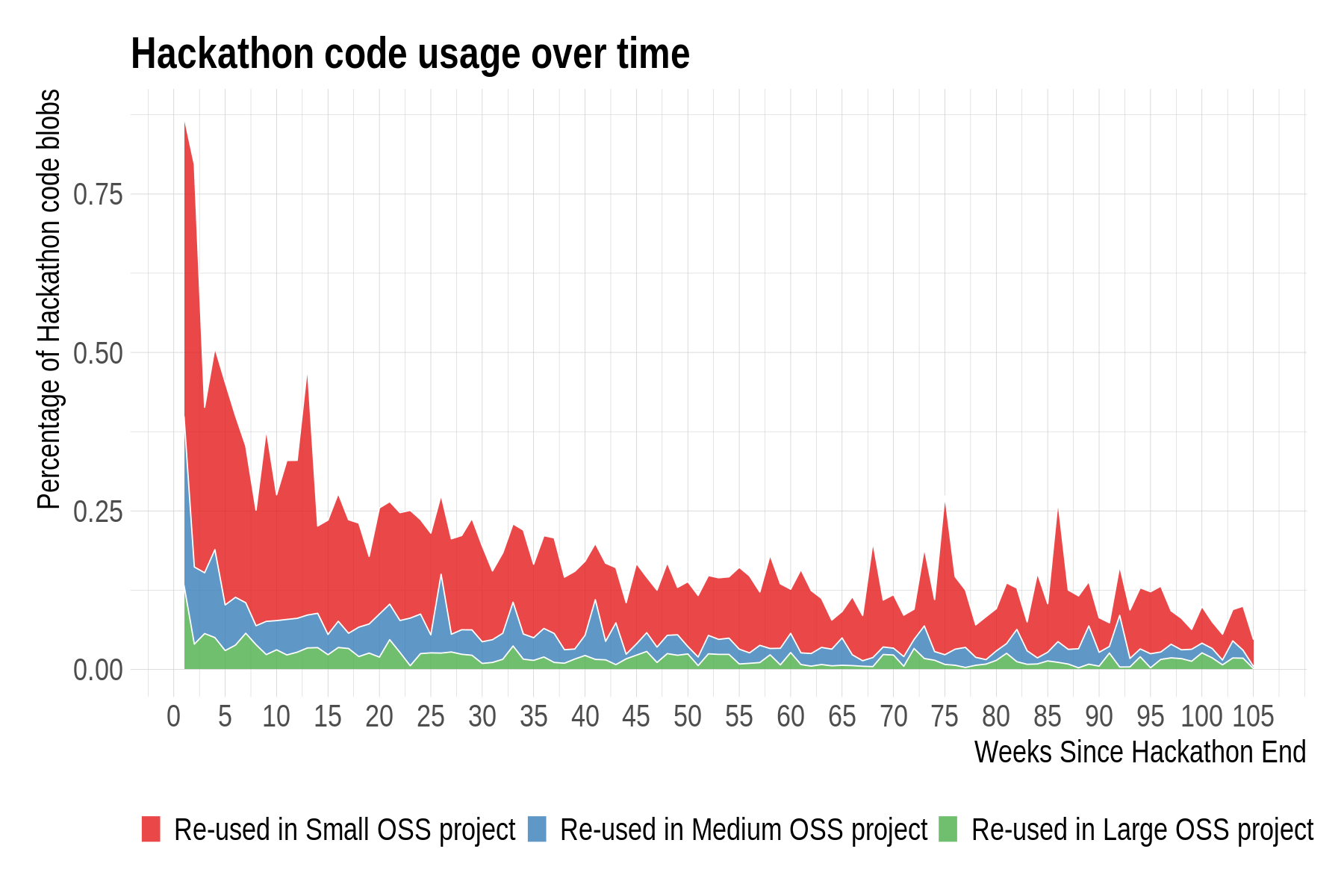}%
%\vsapce{-10pt}
\caption{\textbf{Weekly hackathon code blob reuse in projects of different categories over the period of 2 Years}}
\label{fig:rq2-time}
%\vsapce{-15pt}
\end{figure}

We were interested in exploring the temporal dynamics of code blob reuse as well. Therefore, we looked at the reuse of hackathon code blobs over time for the duration of two years (104.3 weeks) after the corresponding hackathon event ended. The result of that analysis is shown in \cref{fig:rq2-time}, which shows the \textit{weekly} hackathon code blob reuse for 2 years after the end of the corresponding hackathon event, with the fraction of the total number of hackathon code blobs (581,579) reused per week on the Y-axis.
As we can see from this plot, while overall 28.8\% of the hackathon code blobs were reused, over the span of a single week, no more than 0.8\% of the blobs got reused. This finding is in line with prior work on hackathon project continuation (e.g. Nolte et al.~\cite{nolte2020what} found that continuation activity drops quickly within one week after a hackathon before reaching a stable state) within the same repository that the team used during the hackathons. A clear trend of the code blob reuse dropping and then saturating after some time is visible, which is significant because it indicates that the \textit{code created in the hackathon events continue to bear some value even after 2 years} have passed after the event. For code blob reuse in \textit{Small} projects, the knee point comes after around 10-15 weeks, while for the \textit{Medium} and \textit{Large} projects, it comes much earlier, in around a month. It is also a bit surprising to see code blob reuse peaking so soon after the event, but this could be due to the participants of the event putting/influencing people they know to put the code blobs they think are valuable to some other project where they think it might be of use. It might also be because, the chances of a code file being modified increases with time, which results in the SHA1 value being changes, effectively giving rise to a new blob. This distinction has not been studied in prior work on hackathon code.

\hypobox{\textbf{RQ2.c}: Out of the reused hackathon code blobs 57.73\% were used in \textit{Small} projects, 32.85\% in \textit{Medium} projects, and 9.42\% in \textit{Large} projects. Most of the reused blobs were related to web/mobile apps/frameworks. The temporal dynamics of code blob reuse show a clear trend of it reducing over time, and then saturating to a stable value.}

\subsubsection{Do the hackathon code creators foster reuse? (RQ 2.d)}

\begin{figure}
\centering
\includegraphics[width=0.75\linewidth]{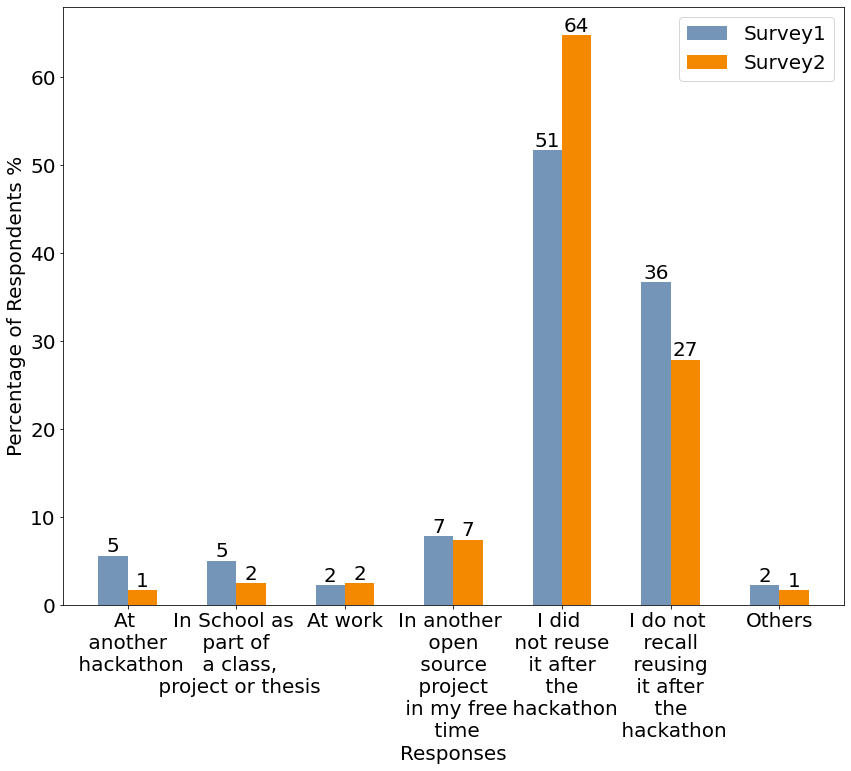}%
%\vsapce{-10pt}
\caption{\textbf{If the code creators reused the code contained in this file after the hackathon}}
\label{fig:rq2d2}
%\vsapce{-15pt}
% \end{subfigure}
% \caption{\textbf{Did the code creators foster reuse of their code by sharing it online (a) or reusing it themselves (b)?}}
\end{figure}
\begin{figure}[!ht]
\centering
% \begin{subfigure}{.48\textwidth}
%\vsapce{-10pt}
\includegraphics[width=0.75\linewidth]{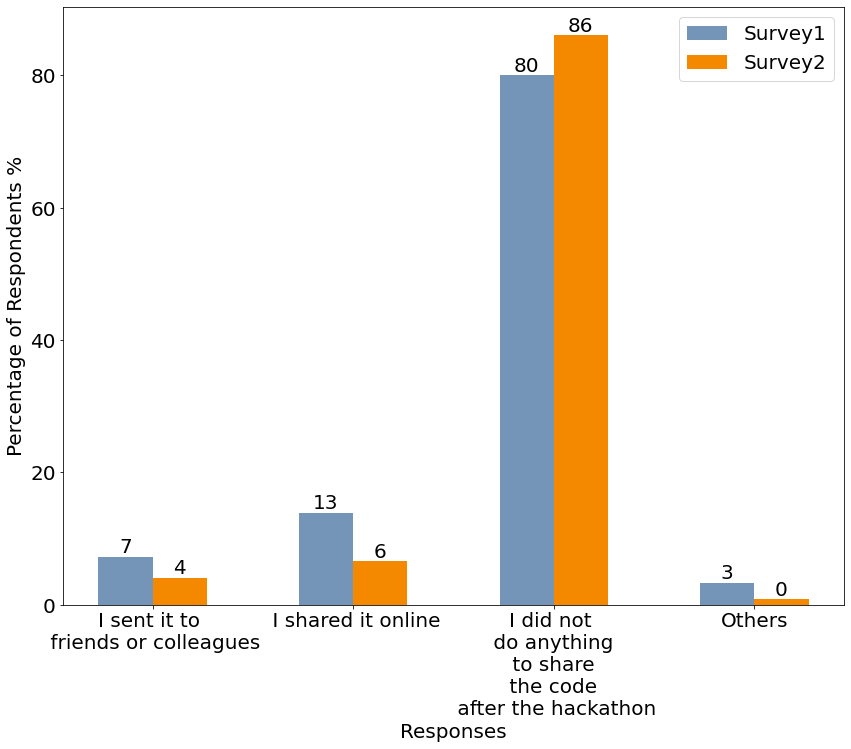}%
%\vsapce{-10pt}
\caption{\textbf{If/where did the code creators share the code contained in this file after the hackathon}}
\label{fig:rq2d}
%\vsapce{-15pt}
% \end{subfigure}
\end{figure}
% \begin{subfigure}{.48\textwidth}

In order to investigate whether the hackathon project members who created a piece of code took any steps to foster its reuse, we focused on two specific aspects - if they themselves reused the code (blob) elsewhere and if they took any steps for sharing the code~\footnote{Although the participants did share their code on GitHub during the hackathons (that is how we found the code), that is not the focus here. We asked the developers: ``What if anything did you do to share the code contained in this file \textbf{after} the hackathon?'' (see ~\cref{sec:app:instruments} for further details) and based on the result from the pilot study, the intent of the question should be clear to the participants that we are not talking about sharing the code in \GH but focusing on if they did something extra after the event rather than just having the repository publicly available.
}. We gathered the responses from survey scenarios 1 and 2 where these two questions were asked to the participants.

Looking at whether the code creators reused the code (blob) themselves (\cref{fig:rq2d2}), we notice that most of the participants did not reuse the code or do not recall whether or not they reused the code though more participants from survey 2 (whose code was not reused as per our collected data) chose the former option. We did notice that a few participants said they reused the code at a different hackathon/at work/at school/in another project and more participants from survey 1 (whose code was reused as per our analysis) indicated that they reused the code. It is worth noting however that some participants from the survey indicated that they reused the code but our analysis indicated that their code wasn't reused. The reason for this discrepancy could be that the code was modified and reused or was not shared openly in GitHub, both of which our analysis does not cover. This aspect is discussed in further detail in \cref{s:limitation}.

In terms of whether the participants shared the code (blob) with others or not (\cref{fig:rq2d}), we see a clear majority of them said they did not share the code though the number is a bit higher for participants of survey 2. Some participants, however, indicated that they shared it with friends/colleagues or online, most of whom were participants from survey 1. 

\begin{table}[ht]
\centering
\caption{Logistic Regression model for statistically explaining code blob reuse by if the authors foster code blob reuse by reusing/sharing the code }\label{t:reuse-authorfoster}
\resizebox{\textwidth}{!}{%
\begin{tabular}{rrrr|rrrr}
\multicolumn{4}{c|}{Code authors reusing the code themselves} & \multicolumn{4}{c}{Code authors sharing the code} \\ \hline
 & Estimate & Std. Error & p-Value &  & Estimate & Std. Error & p-Value \\ \hline
\cellcolor[HTML]{FE0000}Did not reuse & \cellcolor[HTML]{FE0000}-0.06 & \cellcolor[HTML]{FE0000}0.06 & \cellcolor[HTML]{FE0000}0.07 & shared:NO & -0.31 & 0.04 & 2.7e-9 \\
\cellcolor[HTML]{FE0000}Do not recall & \cellcolor[HTML]{FE0000}-0.16 & \cellcolor[HTML]{FE0000}0.06 & \cellcolor[HTML]{FE0000}0.95 &  &  &  &  \\ \hline
\end{tabular}%
}
\end{table}

In order to get a more definitive picture of whether or not the authors' actions on fostering code reuse has a significant impact on the code blobs being reused, we ran two separate logistic regression models in order to control for the other variables. For ease of interpretation, we grouped the answers for whether the authors reused the code themselves into 3 groups - ``Reused'', ``Did not reuse'', and ``Do not recall'', and the answers for if they shared the code after the hackathon into two groups - ``shared:YES'' and ``shared:NO'' by congregating the appropriate responses.
The results, as shown in~\cref{t:reuse-authorfoster}, highlight that whether the authors' reused the code themselves does not have a significant effect on actual code blob reuse  (the ``Reused'' category was the reference category for this situation so it was removed from the results~\footnote{This might need some clarification for the readers not familiar with the underlying statistics. For a categorical variable with multiple categories, most solvers make dummy variables under the hood, choosing one of the categories, usually the first one, as a reference category. The question of the significance is therefore effectively transformed from ``Is there a significant effect of this independent variable on this dependent variable?'' to ``Is there a significant effect of instances belonging to this category compared to instances belonging to the reference category on the dependent variable?''. For binary variables this would be practically identical to the original question, but for variables with more levels/categories, it needs some pondering to properly interpret the result. An example: Imagine the case where you are measuring the amount of sugar in three colors of apples: red, green, and yellow.  Let’s say red is the reference level, so you have two dummy variables:  one for green and one for yellow. The question then becomes: ``Is there an effect of green apples relative to red apples on sugar?   And how about yellow apples relative to red?''\label{fn:stat} }), but whether they shared the code after or not does. This is an important insight, showing the importance of getting the code out to others for fostering the reuse of code.

% Although a statistical analysis found the effect of sharing the code or reusing the code is not significant in explaining whether or not the code was reused as per our analysis, we consistently noticed that participants whose code was reused (survey 1) more commonly shared or reused the code themselves.

\hypobox{\textbf{RQ2.d}: Although most participants of surveys 1 and 2 indicated they did not share or reuse the code, participants from survey 1, whose code we found was reused, did reuse/share the code more often than participants of survey 2. Moreover, the authors' sharing of the code after the hackathon seem to have a significant effect on whether the code blob was reused or not.}

\subsubsection{Are the hackathon code creators aware of their code being reused/feel that their code might be useful to others? (RQ 2.e)}

\begin{figure}[!ht]
% \begin{subfigure}{.5\textwidth}
\centering
\includegraphics[width=0.75\linewidth]{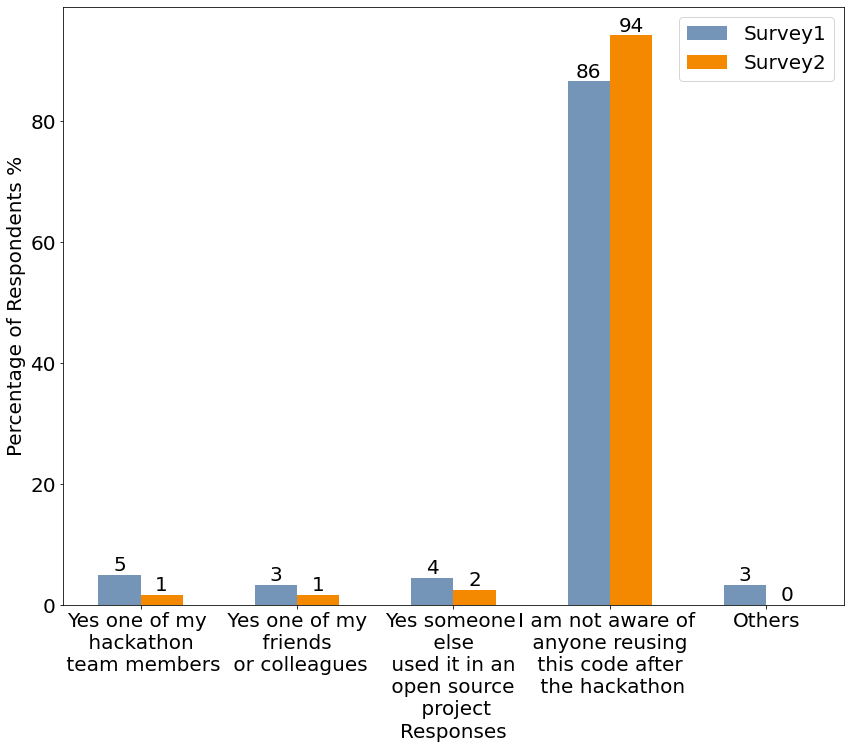}%
%\vsapce{-10pt}
\caption{\textbf{Awareness of others reusing the code}}
\label{fig:rq2e-a}
%\vsapce{-15pt}
% \end{subfigure}
\end{figure}
\begin{figure}
% \begin{subfigure}{.5\textwidth}
\centering
\includegraphics[width=0.75\linewidth]{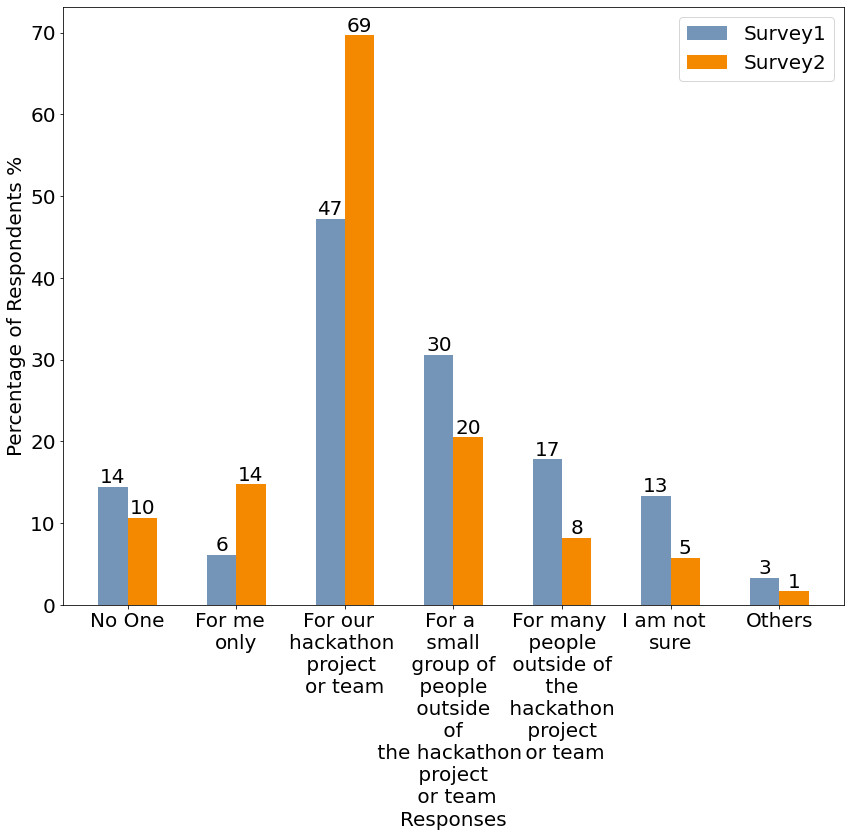}%
%\vsapce{-10pt}
\caption{\textbf{For whom would the code be useful?}}
\label{fig:rq2e-b}
%\vsapce{-15pt}
% \end{subfigure}
% \caption{\textbf{Code creators' awareness of others reusing the code (a) and for whom they think the code might be useful (b)}}
\end{figure}

When we asked the participants of surveys 1 and 2 about whether they are aware that their code is being reused or not, most of them indicated (\cref{fig:rq2e-a}) that they are not aware of it, with more participants from the survey 2 choosing this option. Very few (around 15\%) participants from survey 1, whose code was reused as per our analysis, indicated they know about the reuse, and even fewer (around 5\%) participants from survey 2 chose this option. The reason why we did not capture the reuse indicated by the participants of survey 2 could be the same as what we discussed for the last question.

When asked who the participants think their code might be useful for (\cref{fig:rq2e-b}), most participants said it would only be useful for their project/team. The percentage values in \cref{fig:rq2e-b} sum up to over 100\% because this was a multiple-answer question. Here we notice a very clear distinction between the responses from participants of surveys 1 and 2. Respondents of survey 2 chose that the code would only be useful for them or their team/project much more often than those of survey 1. In contrast, participants from survey 1 chose that the code would be useful for other people as well more often than participants from survey 2, though they also expressed uncertainty about the usefulness of the code or thought it won't be of use for anyone more often. However, the overall result leads us to believe that there is something about the nature of some pieces of code that makes them more likely to be useful for others and, consequently, be reused more often, and at least some of the respondents, whether consciously or subconsciously, are aware of that.

\begin{table}[ht]
\centering
\caption{Logistic Regression model for statistically explaining code blob reuse by the authors' awareness of reuse and feeling about the usefulness of the code}\label{t:reuse-authoraware}
\resizebox{\textwidth}{!}{%

\begin{tabular}{llll|rrrr}
\multicolumn{4}{p{5cm}|}{Code authors' awareness of code reuse} & \multicolumn{4}{p{5cm}}{Who the authors think the code might be useful for} \\ \hline
\multicolumn{1}{r}{} & \multicolumn{1}{r}{Estimate} & \multicolumn{1}{r}{Std. Error} & \multicolumn{1}{r|}{p-Value} &  & Estimate & Std. Error & p-Value \\ \hline
\rowcolor[HTML]{FE0000} 
aware:YES & 0.04 & 0.12 & 0.27 & \cellcolor[HTML]{FE0000}MeOnly & \cellcolor[HTML]{FE0000}-0.17 & \cellcolor[HTML]{FE0000}0.09 & \cellcolor[HTML]{FE0000}0.07 \\
 &  &  &  & OurTeam & -0.17 & 0.07 & 0.01 \\
 &  &  &  & SmallGroupOutside & 0.15 & 0.06 & 0.02 \\
 &  &  &  & \cellcolor[HTML]{FE0000}ManyOutsiders & \cellcolor[HTML]{FE0000}0.16 & \cellcolor[HTML]{FE0000}0.09 & \cellcolor[HTML]{FE0000}0.08 \\
 &  &  &  & \cellcolor[HTML]{FE0000}NotSure & \cellcolor[HTML]{FE0000}0.14 & \cellcolor[HTML]{FE0000}0.10 & \cellcolor[HTML]{FE0000}0.16 \\
 &  &  &  & \cellcolor[HTML]{FE0000}Other & \cellcolor[HTML]{FE0000}0.30 & \cellcolor[HTML]{FE0000}0.12 & \cellcolor[HTML]{FE0000}0.12 \\ \hline
\end{tabular}%
}
\end{table}

In order to gauge the significance of the effects of these two variables on code reuse, we once again  ran two separate logistic regression models in order to control for the other variables. Similar to what we did for \textbf{RQ2.d}, we grouped the answers for if the authors are aware of reuse into two groups: ``aware:Yes'' and ``aware:NO'' by congregating the relevant responses. However, we decided to keep the original categories for who the authors think the code might be useful for because we believed aggregating responses in this case would lead to a significant loss of information. 
The results, as shown in~\cref{t:reuse-authoraware}, showed that the authors' awareness is not significantly associated with code blob reuse, which is not too surprising. For the question of usefulness of the code, the interpretation is slightly less straightforward, as explained in footnote ~\ref{fn:stat}. It effectively shows that participants who chose that their code would be useful for a few outsiders (small group of people outside of the hackathon project/team) and those who chose that their code would only be useful for their team had  significant difference in getting their code blobs reused compared to those who chose that their code would not be useful for anyone with the former having a higher chance of being reused and the latter having a lower chance, while for rest of the participants there was no significant difference.

\hypobox{\textbf{RQ2.e}: Most participants of hackathons were not aware of their code being reused and most felt that their code is useful only for their team/project. However, participants from survey 1 more often indicated that they know of their code being reused and more commonly felt that their code might be useful to others. The authors' awareness of code reuse did not have a significant impact on actual code blob reuse, and participants who chose that their code would be useful for a few outsiders  and those who chose that their code would only be useful for their team had  significant difference in getting their code blobs reused compared to those who chose that their code would not be useful for anyone, while for rest of the participants there was no significant difference.}

\subsubsection{Do the ideas in hackathon projects get reused? (RQ 2.f)}

\begin{table}[!ht]
\caption{\textbf{Survey 1 and 2 responses for if the participants reused the ideas that arose from the hackathon} }
\label{tab:rq2f}
\resizebox{\textwidth}{!}{%
\begin{tabular}{|c|cc|cc|}
\hline
\multirow{2}{*}{} & \multicolumn{2}{c|}{Survey 1} & \multicolumn{2}{c|}{Survey 2} \\ \cline{2-5} 
 & \multicolumn{1}{c|}{Count} & \begin{tabular}[c]{@{}c@{}}Percentage\end{tabular} & \multicolumn{1}{c|}{Count} & \begin{tabular}[c]{@{}c@{}}Percentage\end{tabular} \\ \hline
Yes & \multicolumn{1}{c|}{57} & 32.2\% & \multicolumn{1}{c|}{34} & 28.6\% \\ \hline
No & \multicolumn{1}{c|}{80} & 45.2\% & \multicolumn{1}{c|}{60} & 50.4\% \\ \hline
Not Sure & \multicolumn{1}{c|}{40} & 22.6\% & \multicolumn{1}{c|}{25} & 21\% \\ \hline
\end{tabular}
}
\end{table}

We also decided to examine whether or not the ideas that arose from a particular hackathon were reused or not in an effort to understand the legacy/continued effect a hackathon might have. The result, as presented in~\cref{tab:rq2f}, indicates that around a third of the participants from survey 1 and slightly fewer participants from survey 2 recall an instance where they reused the ideas. Once again, we notice that more participants from survey 2 indicate they do not recall such an instance, so there could be something in the nature of the project/code the participants wrote that somehow makes a hackathon more/less likely to be useful in the future.

\hypobox{\textbf{RQ2.f}: Around a third of the participants from survey 1 recalled an instance when they reused some ideas from the hackathon and, overall, they recalled such an instance more often than participants from survey 2.}

% \hypobox{\textbf{Hackathon code reuse (RQ2):} Around 28.8\% of hackathon code blobs got reused in other projects, with 57.73\% of the code being used in \textit{Small} projects, 32.85\% in \textit{Medium} projects, and 9.42\% in \textit{Large} projects. Most of the reused blobs were related to web/mobile apps/frameworks. The temporal dynamics of code reuse show a clear trend of it reducing over time, and then saturating to a stable value.}

\subsection{Characteristics affecting code blob reuse (RQ3)}
\label{sec:results:rq3}

Our third and final Research Question was about identifying what project characteristics affect (i.e., can be used to explain) code blob reuse and we formed four hypotheses about what factors might be affecting it, which were presented in \cref{sec:rq}. Using the procedure outlined in Section \ref{sss:rq3-data}, we gathered the variables of interest related to the hypotheses, as presented in \cref{t:variables}. As discussed in Section \ref{sss:rq3-analysis}, we decided to use Generalized Additive Models (GAM) to identify the variables that have a significant impact on code blob reuse.

Since we presumed the variables related to \textbf{H1}, \textbf{H2}, and \textbf{H4} would have a linear effect on the probability of code blob reuse for a project, we kept them as linear terms in the model. The variables related to \textbf{H3} were presumed to have a non-linear effect, so they were used as non-linear terms in the model. The formula we used for invoking the GAM model was:\\
\textit{
Y $\sim$ no.Participant + is.colocated + no.Technology + Before + During + LicenseType + s(pctProse) + s(pctData) + s(pctCode) + s(pctMarkup)}

% \usepackage{booktabs}
% \usepackage{graphicx}
\begin{table}[t]
\caption{Effect of Project Characteristics on Hackathon Code Reuse - Results from the Generalized Additive Model.\\
\textbf{Part A.} showing the results for the \textit{linear} terms, with the associated Estimate, Standard Error, and  p-Values.\\
\textbf{Part B.} shows the results for the \textit{non-linear} terms, with the Effective Degrees of Freedom -- ``edf'' -- a measure of the degree of non-linearity, the p-Values, and the partial effects of each variable on the response ( 0: No Effect, Positive Values: Positive effects, Negative Values: Negative Effects).\\
The ``pctData'' variable, found to be ``not significant'', is shown in RED, and the corresponding effect plot is omitted}
\label{t:rq3}
\resizebox{\linewidth}{!}{%
\begin{tabular}{@{}p{4cm}llc@{}}
\toprule
\textbf{A. Linear Variables (Hypothesis)} & \textbf{Estimate} & \textbf{Std. Error} & \textbf{p-value} \\ \midrule
no.Participant (H1) & 0.2040 & 0.0181  & $<$ 0.0001 \\ 
  is.colocated-\textbf{TRUE} (H1) & 0.2448 & 0.1018 & 0.0162 \\ 
  no.Technology (H2) & 0.0258 & 0.0059 & $<$ 0.0001 \\ 
  Before (H2) & 0.0001 & 0.0000 & $<$ 0.0001 \\ 
  During (H2) & 0.0028 & 0.0004 & $<$ 0.0001 \\ 
  LicenseType-\textbf{OSSLICENSE} (H4) & 0.1248 & 0.0522 & 0.0168 \\ 
  \cellcolor[HTML]{FF8888}LicenseType-\textbf{Other} (H4) & \cellcolor[HTML]{FF8888}-0.0652 & \cellcolor[HTML]{FF8888}0.0666 & \cellcolor[HTML]{FF8888}0.3273 \\\midrule\midrule
\textbf{B. Non-Linear Variables (Hypothesis) } & \textbf{edf} & \textbf{p-value} & \textbf{Partial Effect Plot} \\\midrule
pctProse (H3) & 3.698 & 0.0438 & \includegraphics[width=3.5cm]{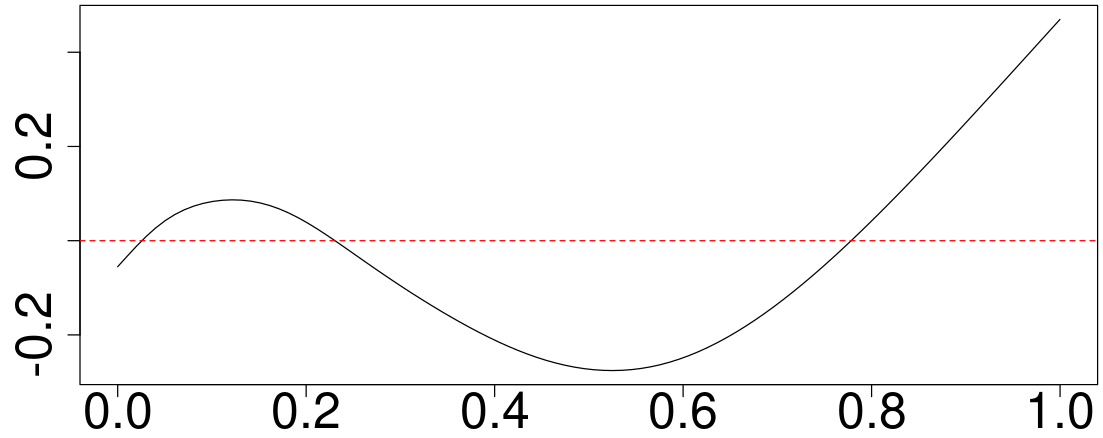} \\\hline
\cellcolor[HTML]{FF8888}pctData (H3) & \cellcolor[HTML]{FF8888}3.716 & \cellcolor[HTML]{FF8888}0.2761 & \cellcolor[HTML]{FF8888} \\\hline
pctCode (H3) & 3.547 & 0.0351 &  \includegraphics[width=3.5cm]{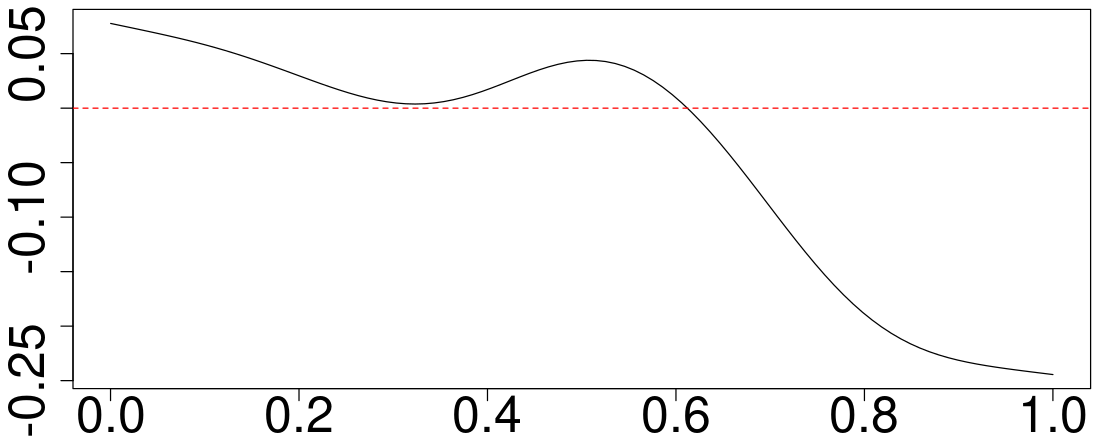} \\\hline
pctMarkup (H3) & 6.779 & $<$0.0001 & \includegraphics[width=3.5cm]{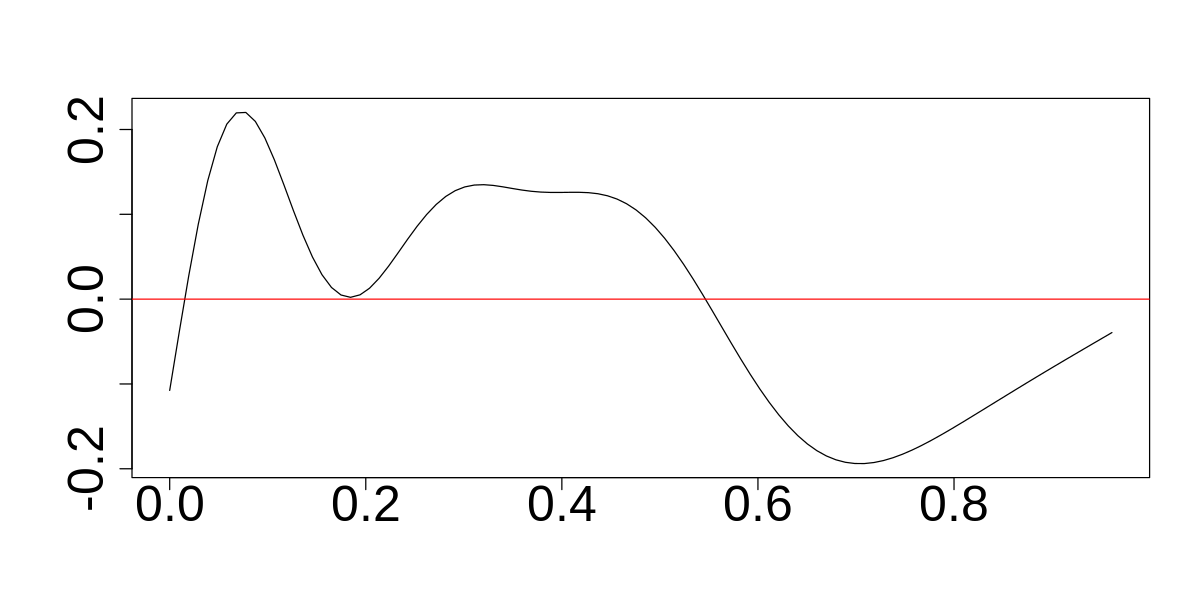} \\ \bottomrule
\end{tabular}%
}
%\vsapce{-20pt}
\end{table}

The result of the analysis is presented in \cref{t:rq3}, which shows that all of the variables related to hypotheses \textbf{H1} and \textbf{H2}, which we assumed would affect code blob reuse, were indeed significant. The effect direction, communicated by the signs of the estimates for the corresponding variables match the rationale we presented in \cref{sec:rq}. These findings are partially in line with prior work on hackathon project continuation in that complex projects for which hackathon participants have prepared prior to the hackathon showed increased continuation activity~\cite{nolte2020what}. In that study, the team size was however negatively related to project continuation while for the study in this paper we found the reverse to be true for code blob reuse. One possible explanation for this discrepancy can be related to larger teams having more opportunities and wider networks to spread the news about their project~\cite{nolte2018you}.

In terms of the usage permission (\textbf{H4}), we found that hackathon projects with an OSS license had a higher chance of having their code blobs reused and the effect was found to be statistically significant as well. However, the effect of a hackathon project having a license of type ``Other'' was statistically insignificant. Regression models do not show the results for all levels for categorical variables like \textit{LicenseType}, but we verified that the effect of a hackathon  project not having a license (\textit{NOLICENSE}) was also statistically insignificant by varying the level order of the \textit{LicenseType} variable in another regression model. 

As for the variables related to \textbf{H3}, the ``pctData'' (see \cref{t:variables} for definition) variable was found to be not significant, but the other three variables were. Since the effects of these variables were non-linear in the model, we decided to observe their partial effect plots, which shows the relationship between the two plotted variables (an outcome and an explanatory variable) while adjusting for interference from other explanatory variables~\cite{pplot}. As demonstrated by the partial effect plots, and the effective degrees of freedom associated with each of these variables, each of them actually have a non-linear effect on the response variable. 

Let's take a closer look at the effects of the three significant variables related to \textbf{H3}. For ``pctProse'',  which refers mostly to documentation files (e.g. Markdown, Text, etc.), we see that having some documentation is good, however, projects which have around half its total content as documentations are not likely to have their code blobs reused. However, rather surprisingly, we see that projects with almost all of their content as documentation are more likely to have their code blobs reused. On closer inspection, we found that these are  quite large projects with a lot of data stored in their corresponding repositories in the form of text or other files (An example of such a project is \url{https://github.com/sreejank/PoliClass}). Therefore, though most of them have a good amount of code, by volume, it appears that it is almost all made up of files of type ``Prose'', which causes this predictor to show positive values for projects very high amount of ``Prose'' files. Such projects however are common in particular in civic and scientific hackathons where participants often develop projects that are related to utilizing specific datasets (e.g.~\cite{nolte2020support,trainer2016hackathon,busby2016closing}). Our finding thus can potentially point to a specific use case that is beneficial for code blob reuse after an event has ended.

On closer inspection of the variable ``pctCode'', we realized that it refers to how much of the total content in a repository is of type ``code'', not the absolute number of code blobs, therefore, having a more balanced repository with a good mix of other types of files signal that it is of higher quality, thus increasing the chance of code blob reuse. This is somewhat expected since for code blobs to be reused it is beneficial to have accompanying documentation as well as use cases (data). As we can observe, projects with over 60\% of their content related to code take a big hit when it comes to their code blobs getting reused. This finding can potentially be due to hackathon teams only having a finite amount of time during an event to actually develop code and the more the code they develop the more likely it is that they do not have time to ``polish" it for reuse.

The behavior of the ``pctMarkup'' variable is more complex than the rest (it also has a higher \textit{edf} value), so it is hard to summarize the interaction without a detailed inspection on a larger dataset, however, it looks like having up to around 60\% markup content (e.g. HTML, CSS, LaTeX, etc.) can lead to a higher propensity of code blob reuse. 

\hypobox{\textbf{Characteristics affecting Code Reuse (RQ3):} The hypotheses presented in \cref{sec:rq} were found to hold. All of the variables related to \textbf{H1} and \textbf{H2} were significant and had effects as anticipated. Examining \textbf{H4} revealed that having an OSS license associated with the hackathon project increased the chance of code blob reuse. The effect of the variables related to \textbf{H3} were more complex, which led to additional insights about hackathon code blob reuse.}

\section{Discussion}
\label{sec:impl}
% Our findings have a number of implications for research and practice. 

Our study reveals the origin story of the code blobs used in hackathon projects in terms of where the code blobs came from, who were the original creators, when it was first created, what are its characteristics in terms of size and programming language, and whether it can be considered a ``template code" or not. 

% \newpage

Before discussing the results and their implications in further detail, we would like to re-emphasize the scope of our study. As mentioned earlier, we are looking at the reuse of code blobs, i.e., when code files are used ``as-is'' without any modifications. This essentially means any type of ``copy-modify-reuse'' scenarios are excluded from our consideration. However, the reuse of code blobs can also be a result of a few different scenarios and an understanding of such scenarios might be useful in better defining the context of our findings. First of all, reusing may be done at the level of complete, ``ready-to-reuse'' components, which would include the ``template code'' we described here. 
The ``template code'' components are generally meant to be reused and such reuse is captured by our study. However, the best practice in reusing a lot of other reusable components usually is to copy the code only in the compilation / deployment directory, not in the source code repository. Therefore, such cases (where the best practices were followed) not captured by the study (both in code reused in hackathons, and in code reused from hackathons which happen to produce ``ready-to-use'' components). But not everybody follows good practices, and repositories can be found with, for example, all dependencies copied to the repository, in which case, our study should capture those.
The second code reuse scenario is when the code that is generally not a ``ready-to-use'' component is copied from one repository and reused in another one, which is not uncommon for OSS development. In this case, our study captures the instances where the code is copied and reused without any modification (even when the file names are different). However, cases where only snippets of code were reused or where the code is reused with any modification would not captured by our study.

We found that of all of the code in the hackathon repositories we investigated, only 9.14\% were created during the hackathon by the number of blobs. The number drops to 8\% if we look at the number of lines of code created. However, it is worth noting that, typically, widely-used frameworks like Django, Rails, jQuery, etc have decades of development history and a large number of contributors. Thus, around 9.14\% of the blobs or 8\% of the lines of code being created during the short duration of hackathon events (72 hours per our assumption) by teams of around 2-3 members (up to a maximum of 10 members --- see \cref{t:variables}), is indeed significant and it highlights the importance of hackathons in generating new code.

The importance of contributions of the hackathon project members in creating/contributing code to the hackathon project repositories was also established by our analysis.
If we consider the code blobs created during and after the hackathon event, as shown by the combined picture in \cref{fig:rq1}, which are the main contributions of the hackathons in terms of code creation, we see that most of that code blobs (97\% for those created during the event, and 93\% for those created after the event) were actually created by the project members. Moreover, we also find that the project members often reuse the code (blobs) they had written earlier in their projects, 15.67\% of all the code blobs belongs to this category. This finding is in line with prior work on hackathon projects in that teams often prepare their projects e.g. by setting up a repository~\cite{nolte2020what} and/or making (detailed) plans on what they want to achieve during an event~\cite{nolte2018you}. If we look at the total lines of code, we see that around 60.5\% of the lines of code in the hackathon repositories were actually contributed to by the project members. Overall, this analysis highlights the effort the hackathon project members put into their projects.

However, in spite of all the effort that goes into developing code for the hackathon projects, hackathons are mostly perceived as one-off events. As we observed from the responses of our surveys, most of the participants of surveys 1 and 2, who were participants in some hackathon in the past as well, indicated that they have little intention of continuing to work on the project even though they were generally satisfied with it (\cref{fig:satInt}). Although a few of the participants later reused the code (\cref{fig:rq2d}) or the ideas from the hackathon (\cref{tab:rq2f}) or even shared some of the code they had created with others/online (\cref{fig:rq2d2}), majority of them did not. All of this seems to indicate that most of the participants do consider hackathons as one-off events and take a ``do it once and forget about it'' approach.

But what about the evidence? Well, the data seem to suggest that around 29\% of the code created during the hackathons, both by the number of blobs and the lines of code, do get shared and around one-tenth of it gets shared in relatively large OSS projects. A number of participants also indicate that they think their code might be useful for others besides their own project (\cref{fig:rq2e-b}). Thus, the evidence seems to indicate that some of the hackathon events can and do have a lasting impact (as we can see from \cref{fig:rq2-time}, reuse of some of the hackathon blobs continue to happen even after 2 years). It is also worth noting that most of the hackathon participants whose code got reused indicated that they were not aware of the reuse (\cref{fig:rq2e-a}), likely for the obvious reason that reuse of open source code is notoriously hard to track without access to a specialized tool like \WOC. Would the participants have felt otherwise about the impact of the work they did during the hackathons if they become aware that their work might be reused? Unfortunately, we do have an answer to that question yet, but it might be worth investigating in the future.

However, the question we do have an answer to is what factors might be affecting code blob reuse and this finding does have some implications for the hackathon project organizers and the participants alike.  They can serve as valuable guidance for scientific and other communities that aim to organize hackathons for expanding their existing code base. Organizers could suggest participating teams attempt projects that do not require developing a large amount of code and rather focus on a specific use case e.g. related to an existing data set. Moreover, they should suggest teams also spend time on not only developing code but also providing additional materials and documentation, and also for the teams to reuse code from their existing projects rather than attempting to develop a lot of original code. This approach can in turn improve efficiency and foster code reuse after an event has ended. The importance of adding an Open Source license was also demonstrated by our analysis, so it might be useful for the participants to add a license to their repository where possible or for the organizers to encourage such a practice.

With respect to research, our findings provide an initial account of how code (with a particular emphasis on code blobs) gets reused and created during a hackathon as well as whether and where it gets reused afterwards. Moreover, they indicate that hackathon code indeed gets reused and that hackathons can thus be more than one-off coding events.

%\vsapce{-5pt}
\section{Limitations and Threats to Validity}
\label{s:limitation}
For our study, we tracked the code generation and usage on a blob level - represented in \WOC by the SHA1 hash value of each blob - which means that we focused only on exact code reuse since any changes in the file contents would lead to a change in the blob SHA1 value that we used to identify each blob. However, it is quite common to make minor changes in a code file while using it in a different context, and that aspect is not captured in our study, nor can we capture the reuse of code snippets.

The \DP dataset does not include the start date of the hackathon events but it is essential information needed to answer our research questions. We assumed the duration of the hackathons to be 72 hours based on existing literature and a manual investigation of 73 randomly selected hackathons 71 of which lasted up to 3 days. However, that might not have been the case for all of the events we studied which may affect the results of RQ1 and RQ3.

We relied on the \GH Linguist tool to categorize files and we only focused on files with type ``Programming", however, the categorization is not infallible, e.g. the type ``Markup" contains HTML and CSS files which could be considered code instead of documentation.

Our license collection and analysis is conducted based on the whole project license used in \GH. However, license information can be included in the files themselves which can be different from the project license, which would mean it would be perfectly clear in certain cases that the file can be reused, even if there is no LICENSE file. Such cases were not captured in our analysis.

We conducted our analysis solely based on the publicly available data. This might sometimes create an issue in the result in terms of the fact that a blob might first be introduced in a private repository and later reused in a hackathon project. However, due to our limitation, we could count this blob as first being created in the hackathon project, which would impact our final results.

While analyzing the results, we used the number of lines of code (LOC) as a measure of code size, which is one of the commonly used measures. However, there might be other ways of measuring code size which we did not consider. Our method of detecting whether a code is ``template code'' i.e. part of some standard library/framework/boilerplate code is based on three simple heuristics. Although we tested the proposed rules with a small-scale ad-hoc test, it is very likely that the proposed method is not entirely accurate and we might have misclassified a number of blobs as a result.

There are also inherent limitations related to the surveys we conducted in addition to the quantitative analysis. First the responses we received reflect a somewhat special sample of participants (those willing to fill a survey) and may differ from the opinions of other participants and may thus be systematically biased. Second, the survey questions are mostly subjective and may have been understood differently from what we have intended. Third, we relied on large parts of the survey on participants selecting options from a list that we compiled. While we made a founded selection of options and provided the possibility to add other options, there might still be other options that the participants thought of but did not mention. Fourth, there is a time difference between the participation of individuals in hackathons or individuals reusing code and taking our survey. We attempted to mitigate this effect by focusing on the most recent hackathons and commits but perceptions still might have changed e.g. between the time a hackathon took place and a participant of that hackathon taking the survey. Fifth, there might be other aspects that affect code reuse that was not in the focus of our survey. Sixth, in selecting the participants we had to focus on individual hackathons, and more precisely, on individual commits. It could be possible that a developer whose one blob was reused after a hackathon might be someone whose code doesn't usually get reused, or vice-versa.  

Finally, in our study we only considered hackathon projects, thus, our findings may not be generalizable to other types of software projects and repositories. Furthermore, our results are solely based on an observational study, not causal analysis, and the results should be interpreted as such.

% We considered each hackathon project as a standalone entity in the study, and we didn't track code usage and spread between different hackathons and hackathon projects. However, it can be common that a developer participate in different hackathons and utilize the same code he developed earlier in each hackathon project. --- This is more of future work, and out of scope, I don't believe it affects the results we found.

% In our analysis, we tracked the code usage and spread based on a classification of the project size where the blobs ended up in. The classification was done based on the number of authors contributing to that project and the number of stars of that project, we didn't analyze if any of the hackathon project members was part of that project or not, which can help us in understanding the popularity of the blobs if it is reused in another projects where the hackathon team was not part of or only used in projects where any of the hackathon team contributed. --- more of future work than limitation.

% We also didn't track how the hackathon project was created, if it is a forked project or a standalone project. However, it may be a common practice for a developer to fork an existing project when he participate in a new hackathon and utilize his existing code. --- How could that compromise the validity of our findings?

%\vsapce{-5pt}
\section{Conclusion and Future Work}
In this study, we investigated the origins of hackathon code (with a particular emphasis on code blobs) and its reuse after an event. We found that most hackathon projects reuse existing code (blobs) and that code (blobs) created during the events also gets reused by other OSS projects later on. Our study also revealed a number of project characteristics that might affect code blob reuse. The majority of the participants from the surveys we conducted indicated that they mostly view hackathons as one-off events and do not do much to foster reuse of the code they created. However, some of them felt that their code might be useful for other people beyond their own project members. 
In summary, our findings agree with most earlier studies and indicate that most of the participants do see hackathons as one-off events. At the same time, it shows some of the hackathon code blobs do get reused, sometimes even after 2 years after the event, reiterating the actual impact of hackathon events, simultaneously providing an account of code blob reuse in Open Source Software.

There are several ways to extend this research, e.g. considering code clones/ snippets while looking for code reuse (e.g. by looking at the associated CTAG tokens - a dataset available in \WOC), identifying other factors that affect code reuse, including code quality~\cite{dey2018usageQuality,Dey2020qualityEMSE,dey2020modeling}, project popularity~\cite{dey2019patterns}, the developers' mastery on the project topics~\cite{dey2021representation}, the supply chain of a particular software~\cite{dey2018dependency,amreen2019methodology} etc., and if copying code might have any effect on a developer's pull request being accepted~\cite{dey2020effect}. Looking deeper into the code created during the hackathons, it might also be interesting to see to what extent the teams use bots~\cite{dey2020botdetection,dey2020botse} which might aid in the understanding of hackathon code reuse as well.
We hope that further studies will explore these and other related topics, and give us a clearer understanding of the impact of hackathons and code reuse.

\begin{acknowledgements}
The work was supported, in part, by Science Foundation Ireland grant 13/RC/2094\_P2 and by NSF awards 1633437, 1901102, and 1925615.
\end{acknowledgements}

\bibliographystyle{spbasic} 
\bibliography{references}

\clearpage 
\appendix

\section{Appendix}
\label{sec:app}

\subsection{Survey instruments}
\label{sec:app:instruments}
% \begin{table}[h]
% \begin{center}

\begin{longtable}{|p{\textwidth}|}
\caption{\textbf{Survey 1 sent to individuals who had created a blob during a hackathon that was reused after the event and to individuals who had created a blob during a hackathon that was not reused.}}
\label{tab:app:instruments:survey1}\\
\hline
\rowcolor{Gray}
Perception about the code contained in the blob \\
\hline \\ 
The commit was made by GITHUB\_HANDLE. Is this your handle? (yes, no, not sure) \\
Where to the best of your knowledge did the code contained in this file originate from? Please select all options that apply. \\
\tableIndent(1) I wrote it during the hackathon. \\
\tableIndent(2) I reused code that I had written before the hackathon. \\
\tableIndent(3) My team members wrote it during the hackathon. \\
\tableIndent(4) My team members reused code they had written before the hackathon. \\
\tableIndent(5) From friends or colleagues. \\
\tableIndent(6) From other GitHub repositories. \\
\tableIndent(7) From the web (e.g. on Stackoverflow or forums). \\
\tableIndent(8) It was generated by a tool. \\
\tableIndent(9) I am not sure. \\
\tableIndent(10) Other: \\
Where, if at all, did you REUSE the code contained in this file after the hackathon? Please select all options that apply. \\
\tableIndent(1) At another hackathon \\
\tableIndent(2) In School as part of a class, project or thesis \\
\tableIndent(3) At work \\
\tableIndent(4) In another open source project in my free time \\
\tableIndent(5) I did not reuse it after the hackathon \\
\tableIndent(6) I do not recall reusing it after the hackathon \\
\tableIndent(7) Other: \\
What if anything did you do to share the code contained in this file after the hackathon?  Please select all options that apply. \\
\tableIndent(1) I sent it to friends or colleagues \\
\tableIndent(2) I shared it online (e.g. via GitHub, Stackoverflow, Social media) \\
\tableIndent(3) I did not do anything to share the code after the hackathon \\
\tableIndent(4) Other: \\
For whom do you think this file might be useful? Please select all options that apply. \\
\tableIndent(1) Only for me \\
\tableIndent(2) For our hackathon project or team \\
\tableIndent(3) For a small group of people \\
\tableIndent(4) For many people \\
Are you aware of anyone else REUSING the code contained in this file after the hackathon? Please select all options that apply. \\
\tableIndent(1) Yes one of my hackathon team members \\
\tableIndent(2) Yes one of my friends or colleagues \\
\tableIndent(3) Yes someone else used it in an open source project \\
\tableIndent(4) I am not aware of anyone reusing this code after the hackathon \\
\tableIndent(5) Other: \\ [0.8ex]
\hline
\rowcolor{Gray}
Perception about the hackathon project \\
\hline \\ [-2.2ex]
Can you recall any instance where you reused ideas that arose from this hackathon after it had ended? (yes, no, not sure) \\ [0.8ex]
\hline
\rowcolor{Gray}
Perceived usefulness of the hackathons project (based on \cite{reinig2003toward,filippova2017diversity}), anchored between strongly disagree and strongly agree \\
\hline \\ [-2.2ex]
I am satisfied with the work completed in this team. \\
I am satisfied with the quality of my team's output. \\
My ideal outcome coming into my team was achieved. \\
My expectations towards my team were met. \\
We lacked important skills to complete our project. \\ [0.8ex]
\hline
\rowcolor{Gray}
Intentions to continue working on the hackathon project (based on \cite{bhattacherjee2001understanding}), anchored between strongly disagree and strongly agree \\
\hline \\ [-2.2ex]
I intend to continue working on this project rather than not continue working on it. \\
My intentions are to continue working on this project rather than any other project. \\
If I could, I would like to continue working on this project as much as possible. \\ [0.8ex]
\hline
\rowcolor{Gray}
Demographics \\
\hline \\ [-2.2ex]
How old are you currently? (18 to 24, 25 to 34, 35 to 44, 45 to 54, 55 to 64, 65 to 74, 75 or older, Prefer not to say) \\
Are you...? (Female, Male, Non-binary, Prefer not to say) \\
Do you consider yourself a minority? (For example in terms of race, gender, expertise or in another way)? (Yes, No, Prefer not to say) \\
About how many years of experience do you have as a contributor to open source projects in general? (1, 2, 3, 4, 5+) \\ [0.8ex]
\hline
\end{longtable}
% \end{center}

% \vspace{-10pt}
% \end{table}

% ################################################################

\begin{table}[h]
\caption{\textbf{Survey 2 sent to individuals who had reused a blob that was created during a hackathon.}}
\label{tab:app:instruments:survey2}
\begin{center}
\begin{tabular}{|L{\textwidth}|}
\hline
\rowcolor{Gray}
Perception about the code contained in the blob \\
\hline \\ [-2.2ex]
The commit was made by GITHUB\_HANDLE. Is this your handle? (yes, no, not sure) \\
Where to the best of your knowledge did the code contained in this file originate from? Please select all options that apply. \\
\tableIndent(1) I wrote it. \\
\tableIndent(2) From friends or colleagues. \\
\tableIndent(3) From other GitHub repositories. \\
\tableIndent(4) From the web (e.g. on Stackoverflow or forums). \\
\tableIndent(5) It was generated by a tool. \\
\tableIndent(6) I am not sure. \\
\tableIndent(7) Other: \\ [0.8ex]
\hline
\rowcolor{Gray}
Perceived ease of use (based on \cite{davis1989perceived}), anchored between strongly disagree and strongly agree \\
\hline \\ [-2.2ex]
Learning to use the code in this file was easy for me. \\
I found it easy to get the code in this file to do what I want it to do. \\
It was easy for me to become skillful at using the code in this file. \\
I found the code in this file easy to use. \\ [0.8ex]
\hline
\rowcolor{Gray}
Demographics \\
\hline \\ [-2.2ex]
How old are you currently? (18 to 24, 25 to 34, 35 to 44, 45 to 54, 55 to 64, 65 to 74, 75 or older, Prefer not to say) \\
Are you...? (Female, Male, Non-binary, Prefer not to say) \\
Do you consider yourself a minority? (For example in terms of race, gender, expertise or in another way)? (Yes, No, Prefer not to say) \\
About how many years of experience do you have as a contributor to open source projects in general? (1, 2, 3, 4, 5+) \\ [0.8ex]
\hline
\end{tabular}
\end{center}
\end{table}

\end{document}